\documentclass[runningheads]{llncs}
\usepackage{geometry}
\usepackage{url}

\usepackage{breakurl}

\usepackage{subcaption}
\usepackage{xcolor}
\usepackage{hhline}
\usepackage{multirow}
\usepackage{makecell}
\usepackage{eurosym}
\usepackage{amsfonts}
   
\usepackage{amssymb}
\usepackage{tabularx}
\usepackage{placeins}
\usepackage{color}
\usepackage[ruled,vlined]{algorithm2e}
\usepackage{algorithmicx}
\usepackage{mathrsfs}
\usepackage{algpseudocode}
\usepackage{pifont} 
\newcommand{\cmark}{\ding{51}}%
\newcommand{\xmark}{\ding{55}}%

\SetCommentSty{mycommfont}
\usepackage{setspace}

\usepackage{multirow}
\usepackage{graphicx}

\begin{document}
	\title{Precision Health Data: Requirements, Challenges and Existing Techniques for Data Security and Privacy}

	\author{Chandra Thapa \and Seyit Camtepe}
	\institute{CSIRO Data61, Australia\\
		\email{\{chandra.thapa, seyit.camtepe\}@data61.csiro.au}
	}

\authorrunning{Thapa et al.}
	\maketitle


	
	
	

	

	\begin{abstract}
		Precision health leverages information from various sources, including omics, lifestyle, environment, social media, medical records, and medical insurance claims to enable personalized care, prevent and predict illness, and precise treatments. It extensively uses sensing technologies (e.g., electronic health monitoring devices), computations (e.g., machine learning), and communication (e.g., interaction between the health data centers). 
		As health data contain sensitive private information, including the identity of patient and carer and medical conditions of the patient, proper care is required at all times. Leakage of these private information affects the personal life, including bullying, high insurance premium, and loss of job due to the medical history. Thus, the security, privacy of and trust on the information are of utmost importance.
		Moreover, government legislation and ethics committees demand the security and privacy of healthcare data.
		Besides, the public, who is the data source, always expects the security, privacy, and trust of their data. Otherwise, they can avoid contributing their data to the precision health system. Consequently, as the public is the targeted beneficiary of the system, the effectiveness of precision health diminishes.
		Herein, in the light of precision health data security, privacy, ethical and regulatory requirements, finding the best methods and techniques for the utilization of the health data, and thus precision health is essential. 
		In this regard, firstly, this paper explores the regulations, ethical guidelines around the world, and domain-specific needs. Then it presents the requirements and investigates the associated challenges.
		Secondly, this paper investigates secure and privacy-preserving machine learning methods suitable for the computation of precision health data along with their usage in relevant health projects. Finally, it illustrates the best available techniques for precision health data security and privacy with a conceptual system model that enables compliance, ethics clearance, consent management, medical innovations, and developments in the health domain.
	\end{abstract}
	
	
%
	

	\section{Introduction}
	
\begin{figure}[!t]
		\centering
		\includegraphics[width=0.55\linewidth, trim=1cm 1cm 1cm 0.5cm, clip=true]{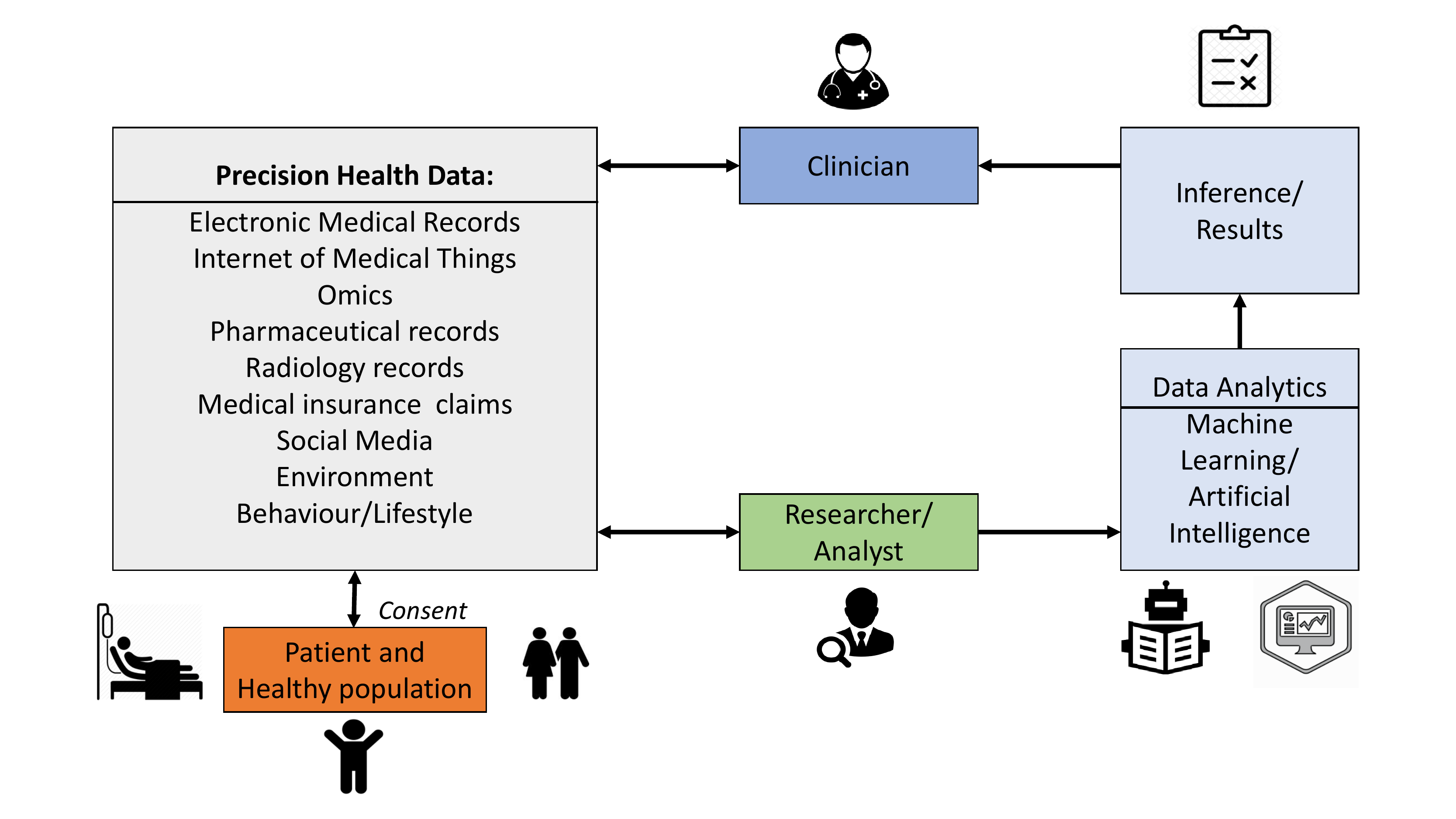}
		\caption{Precision health ecosystem} 
		\label{fig:phecosystem}
	\end{figure}

	\color{black}
	Precision health is a precise, personalized, prescriptive, and preventive approach to healthcare. As illustrated in Figure~\ref{fig:phecosystem}, it leverages collective information from diverse sources, including omics (e.g., genomics), lifestyle, environment, social media, internet of medical things, medical history, pharmaceuticals, and medical insurance claims~\cite{precisionhealth2,precisionhealth3}. Precision health will not only refine the current health care practices of providing care after an illness, but also predict, prescribe, and prevent the illness before they develop. For example, the risk of type 2 diabetes mellitus is identified through longitudinal study (8 years) of the clinical measures and tests, including omics profiling, microbiome, and wearable monitoring~\cite{naturenew}. In another work, online review data of restaurants on social media are leveraged to predict the hygiene of the restaurant and health risks~\cite{restaurantdata}. People can take advantage of these predictions at the right time to avoid potential health risks. Besides, the preventive approach (e.g., detection and treatment of illness at early stages) and precision diagnosis (e.g., right drugs and correct diagnosis) in precision health enables a reduction in the healthcare cost, which is expensive and ever increasing~\cite{ref6}. For example, the USA spent \$3.6 trillion in 2017, which is 4.4\% higher than in 2017~\cite{ushealthspending}. Similarly, Australia spent \$181 billion on health care in 2016-17, which is 1.6\% higher than the average over 2011-2015\cite{ref2}. 
	
	The main fuel of precision health for its operation is health data, which is growing at a fast pace. The growth is due to electronic health records (EHR), medical images, and the internet of medical things (IoMT), including wearable devices (e.g., fitness trackers such as FitBits). It is estimated that 2,314 exabytes of health data will be produced in 2020~\cite{ref5}. These data are profiling individuals, and it can be leveraged by clinicians and researchers in the precision health ecosystem. Usually, these data are decentralized in nature and non-iid in characteristics. 
	
	The precision health data has mainly seven stages in its life-cycle, namely data generation, collection, processing (e.g., health data cleaning and data encryption), storage, management (e.g., creating metadata and access control), analytics and inference. Data analytics is an integral part of precision health. It is a systematic use of data combined with quantitative as well as qualitative analysis to make decisions~\cite{ref7}. It supplies techniques to transform accumulated raw data into valuable insights, and their utilization enables an evidence-based healthcare delivery. For example, mutation prediction~\cite{ref15}. Artificial Intelligence (AI) and Machine learning (ML) boost analytic, and health data analytics have been a part of healthcare~\cite{ref9,ref10}. The analytic has a huge impact on medical research, daily life, patient experience, ongoing care, prediction, and prevention~\cite{ref18,naturelungcancer}. Besides, it is a growing industry~\cite{ref8}. It saves a considerable amount of expenditure in the healthcare economy. It is estimated that the key clinical health AI applications can save \$150 billion annual savings for the USA healthcare economy by 2026~\cite{ref21}.

	Based on the precision health data life-cycle, we can broadly divide health data into three categories, namely data-at-rest (stored, not currently transmitted or processed), data-in-transit (data currently being transferred from one part to another) and data-in-use (data in memory, including CPU caches and registers). Data analytics predominantly deals with data-in-use.
	Although precision health has the potential to revolutionize current healthcare, it faces difficulties due to security, privacy, ethical and legal concerns related to health data of all categories. It contains sensitive information of patient and carer, including their identity, medical condition of the patient, and cure. A proper consent needs to be taken for the use and reuse of the data from their owner. In addition, care must be taken during all stages of the data life cycle because leakage of this information affects personal life, including poor social networking and loss of job due to the medical history, no employment, and high insurance premium. Patient engagement survey 2018 by MedicalDirector in partnership with HotDoc~\cite{ref3} found that 93\% and 91\% Australian rate security and privacy, respectively, are a top concern. This concern is not limited to Australians; it is everywhere. Besides, among all data breaches, health data covers a significant portion, and it is increasing~\cite{ref13,ref14,ref19,databreach2019}. Governments are also concerned about the security and privacy of health data. They have been regulating and managing the concerns through government policies and legislation (refer to Section~\ref{sec:law} for details).

	We refer the sensitive information to protected health information or personal information, which are defined in the following. These terms refer to the same information in the health domain, so we refer to them by simply personal information (PI) in the remainder of this paper for convenience. \\
	\emph{Protected health information}: According to the health insurance portability and accountability act (HIPPA) of USA, protected health information (PHI) includes all individually identifiable information, including demographic data, medical histories, test results, insurance information, other information used to identify a patient or provide healthcare services or healthcare coverage~\cite{ref16}.\\
	\emph{Personal information}: Based on Privacy Act 1988, Australia, personal information (PI) is information or an opinion that identifies you or could identify you and includes information about your health~\cite{ref17}. In the European Union's general data protection regulation (GDPR)~\cite{gdpr}, personal information is defined as any data that relates to an identified or identifiable individual. This data includes online identifiers (e.g., IP address), sales databases, location information, CCTV footage, bio-metric data, loyalty scheme records, and health information.
	
	%
	%
	%
	
	The current legal and ethical aspects are important for health data, which includes PI. For precision health data security, privacy and trust, there is no elaborative work that investigates these aspects to identify the requirements and challenges along with recently evolving enabler techniques, including privacy-preserving distributed collaborative machine learning techniques. The requirements guide to maintain compliance, and the techniques ensure it in precision health. Refer to Table~\ref{tab:survey} for related works.  
	\begin{table}[!h]
		\centering
		\footnotesize
		\caption{Related works in health data security and privacy}
		\renewcommand{\arraystretch}{1.1}
		\resizebox{\textwidth}{!}{%
			\begin{tabular}{|c|l|l|l|}
				\hline
				Reference & Focus                                                                          & Review                                                                                                                                                                                                                                                                                                                                                                         & Technology focus                                                                                                                                                                                                                                                                                                             \\ \hline
				\cite{survey1} & \begin{tabular}[c]{@{}l@{}}Electronic health record\\ system\end{tabular}      & \begin{tabular}[c]{@{}l@{}}Compliance, information systems acquisition\\ development, access control, communications\\ and operations management, human resource\\ security\end{tabular}                                                                                                                                                                                   & \begin{tabular}[c]{@{}l@{}}Pseudo-anonymity technique, encryption,\\ authentication systems, audit logs\end{tabular}                                                                                                                                                                                                     \\ \hline
				\cite{survey2} & e-Health clouds                                                                & \begin{tabular}[c]{@{}l@{}}Requirements for e-health cloud, \\ privacy-preserving approaches\end{tabular}                                                                                                                                                                                                                                                                      & \begin{tabular}[c]{@{}l@{}}Various cryptography based approaches, \\ including homomorphic encryption and \\ searchable encryption\end{tabular}                                                                                                                                                                              \\ \hline
				\cite{survey3} & Big data in healthcare                                                         & \begin{tabular}[c]{@{}l@{}}Data protection law of seven countries (e.g.,\\ HIPAA of USA and Data Protection\\ Directives of EU)\end{tabular}                                                                                                                                                                                                                                   & \begin{tabular}[c]{@{}l@{}}Authentication, encryption,\\ de-identification, access control\end{tabular}                                                                                                                                                                                                                      \\ \hline
				\cite{survey4} & Biomed data science                                                            & \begin{tabular}[c]{@{}l@{}}Brief discussion on HIPAA, research ethics,\\ and patient's viewpoint\end{tabular}                                                                                                                                                                                                                                                              & \begin{tabular}[c]{@{}l@{}}De-identification methods, data\\ anonymization methods, privacy-\\preserving predictive modeling\\  (use cases of federated learning and \\ encrypted data analysis)\end{tabular}                                                                                                               \\ \hline
				\cite{survey5} & Big healthcare data                                                            & \begin{tabular}[c]{@{}l@{}}Briefly mentioned salient features of data\\ protection laws of nine countries (e.g., HIPAA\\ USA, Data projection directive EU)\end{tabular}                                                                                                                                                                                                    & \begin{tabular}[c]{@{}l@{}}Authentication, encryption, data masking,\\ access control, monitoring and auditing, \\ de-indentification, HybrEx\end{tabular}                                                                                                                                                               \\ \hline
				\cite{survey6} & eHealth cloud security                                                         & \begin{tabular}[c]{@{}l@{}}US standards (e.g., HIPAA), International\\ Standards (e.g., ISO/IEC 27000-series), \\ GDPR EU\end{tabular}                                                                                                                                                                                                                                      & \begin{tabular}[c]{@{}l@{}}Patient-centric approach, encryption\end{tabular}                                                                                                                                                                                                                                              \\ \hline
				\cite{survey7} & Federated learning                                                             & \begin{tabular}[c]{@{}l@{}}Current progress on federated learning\\ (technical) attacks and defenses\end{tabular}                                                                                                                                                                                                                                                              & \begin{tabular}[c]{@{}l@{}}Federated learning, secure multi-party\\ computation, differential privacy\end{tabular}                                                                                                                                                                                                       \\ \hline
				\cite{survey8} & Human genomic data                                                             & \begin{tabular}[c]{@{}l@{}}Privacy/security problems related to genomic\\ data sharing/computation\end{tabular}                                                                                                                                                                                                                                                               & \begin{tabular}[c]{@{}l@{}}Homomorphic encryption, Intel SGX,\\ differential privacy\end{tabular}                                                                                                                                                                                                                           \\ \hline
				\cite{survey9} & Healthcare                                                                     & \begin{tabular}[c]{@{}l@{}}Challenges for the machine learning techniques,\\ including vulnerabilities in machine learning\\ pipeline, model training, adversarial machine\\ learning and privacy-preserving machine\\ learning (e.g., federated learning in brief),\\ countermeasures against adversarial attacks\\ on models, responsible machine learning\end{tabular} & \begin{tabular}[c]{@{}l@{}}Machine learning applications on\\ prognosis, diagnosis, treatment\end{tabular}                                                                                                                                                                                                                   \\ \hline
				\cite{survey10} & Medical imaging data                                                           & \begin{tabular}[c]{@{}l@{}}Overview of methods for federated, secure and\\ privacy-preserving artificial intelligence\end{tabular}                                                                                                                                                                                                                                         & \begin{tabular}[c]{@{}l@{}}Re-identification issue, federated machine\\ learning, differential privacy, homomorphic \\ encryption, secure multi-party computation,\\ secure hardware implementation (in brief)\end{tabular}                                                                                      \\ \hline
				\cite{survey11} & \begin{tabular}[c]{@{}l@{}}Distributed learning in\\ health care\end{tabular} & Legal context in brief                                                                                                                                                                                                                                                                                                                                                         & \begin{tabular}[c]{@{}l@{}}Machine learning techniques, distributed\\ (deep) learning, block-chain\end{tabular}                                                                                                                                                                                                             \\ \hline
				This paper & \begin{tabular}[c]{@{}l@{}}Precision Health data\end{tabular}               & \begin{tabular}[c]{@{}l@{}} Detailed requirements based on regulations\\  (e.g., GDPR EU), ethical guidelines, and\\ health domain, techniques for (health) data\\ security and privacy, and its consideration\\ in notable health projects\end{tabular}                                                                                                                                                                                                                                                          & \begin{tabular}[c]{@{}l@{}} Machine learning in healthcare, No-peek\\ learning, including federated and split\\ learning on medical informatics, data-centric\\ solutions, cryptography, access control,\\ anonymization, pseudonymization, hardware\\ based security and privacy, homomorphic\\ encryption, multiparty computation,\\ differential privacy\end{tabular} \\ \hline
			\end{tabular}%
		}
		\label{tab:survey}
	\end{table}
	
	\subsection{Our contributions}
	The precision health (PH) data is usually isolated and distributed (e.g., data stored at different hospitals), and it comes from a diverse field. In light of barriers, as mentioned earlier, including security and privacy, it is important to explore the best ways for health data handling and use, including breaking the precision health data silos required to leverage AI/ML efficiently. In this regard, this paper thoroughly explores the requirements, lists out the challenges, and presents the potential candidate methods that enable data privacy and security. 
	Firstly, this paper, in Section~\ref{sec:trustworthyPH}, surveys the data regulations and ethical guidelines from data security and privacy perspectives. This provides detailed requirements for compliance whilst handling PH data. Then, considering the sensitivity of PH data in health decision making, this paper studies the requirements for data trustworthiness. Afterward, based on these requirements, it highlights the existing challenges related to PH data in Section~\ref{sec:challenges}. Secondly, in Section~\ref{sppreservingtech}, it presents current techniques for PH data security and privacy. As the computing environment may not be a trusted platform, PH data privacy and security whilst computation need to be addressed. So, this paper presents the machine learning paradigms and healthcare, including the state-of-the-art privacy-by-design machine learning approaches that ensure PH data privacy and security, in Section~\ref{sec:mlparadigmsandhealthcare}. Together with the relevant health projects, and their PH data security and privacy techniques in Section~\ref{sec:relatedhealthprojects}, the candidate techniques are illustrated with a conceptual system model for the precision health platform to provide an overall picture in Section~\ref{sec:proposedmodel}.

	\section{Requirements for precision health data privacy, security and trust} 
	\label{sec:trustworthyPH}
	
	Precision healthcare is a data-driven healthcare approach. Thus compliance, both to law and ethics, while handling the health data is of utmost importance to avoid penalties and maintain trustworthiness. The proper requirements for privacy, security, and trust of the precision health (PH) data enable us to design and maintain the compliance-friendly techniques and trustworthy platform to handle the PH data. In this regard, we extract the requirements due to law and ethics in the following sections. 
	
	Firstly, we revisit the general definition of and the distinction between law and ethics. According to the Oxford dictionary, the law is the system of rules which a particular country or community recognizes as regulating the actions of its members and which it may enforce by the imposition of penalties~\cite{law}. Law has a set of rules and regulations with legal binding. A government governs it. On the other hand, ethics are moral principles that govern a person's behavior or the conducting of activity~\cite{ethics}. Ethics has a set of guidelines (e.g., code of conduct) governed by individuals, legal and professional norms. It guides us on good and evil, or right and wrong in all aspects of human affairs. Violations of ethical standards result in penalties, including job termination, monetary fines, and legal actions. Ethics and law are complementary to each other, and both are required for better judgment and decision. 
	
	In order to provide a comprehensive requirement as much as possible, the literature is searched and filtered based on its contents and source. For laws (e.g., federal laws) and ethics, relevant documents from governments and institutions are selected that are inclusive and can provide an overall overview. Search is done in the various search engines, including \emph{Google Scholar}, \emph{Scopus}, and \emph{PubMed}, by using keywords, including privacy, security, trust in health data, ethics, and ethical requirements. Many sources have overlapping issues and requirements. Thus, we consider only those which cover most of them. Also, review papers discussing the major laws and ethical requirements for the related field are considered.

	\subsection{Requirements due to law}
	\label{sec:law}
	There have been significant initiations from several countries and organizations towards the acts/regulations of health data privacy, security, and trust.
	These are introduced to ensure the privacy of personal information (PI), which is more relevant when it comes to PH data. To understand the current legislation that provides baseline security and privacy rules around the world, we state the regulations in some countries where EHR is commonly used as an illustration. This includes the USA, EU, and Australia. Broad coverage of regulations around the world is not the scope of this paper. Even within one country, their states can have their separate privacy legislation. For example, states have different general privacy legislation in Australia~\cite{australianstateprivacylaw}, and similarly in the USA, e.g., California Consumer Privacy Act~\cite{ccpa}. However, the requirements are standard and similar.  
	
	\textbf{HIPAA:}
	The USA has a health insurance portability and accountability act (HIPAA) enacted in 1996 for the privacy and security of healthcare information, including health data. The privacy rule standards of HIPAA address the use and disclosure of health data with PI. 
	It assures the protection of PI while allowing the flow of health information (e.g., electronic exchange), which is highly required for medical decisions and well-being~\cite{hipaaprivacy}. 
	HIPAA privacy rules apply to health plans\footnote{Health plans are individuals or group plans that provide or pay the cost of medical care. For example, health insurers, Medicare, health maintenance organizations, and long-term care insurers~\cite{hipaaprivacy}.}, health care clearinghouses\footnote{Health care clearinghouses are entities that process nonstandard information they receive from another entity into a standard or vice versa. For example, billing services, repricing companies, and community health management information systems~\cite{hipaaprivacy}.}, and to health care providers (e.g., hospitals, physicians, dentists, and other practitioners) who transmits health information. 
	HIPAA does not apply to de-identified data, which refers to the data set of the individual from which the PI cannot be traced even by linking with other available data sets. 
	For secondary use of data including research, and analytics, it is mandatory to obtain written authorization from the patients. 
	
	The security standards of HIPAA address the protection of health information that is held or transferred in electronic form. The main aim of the standards is to protect the privacy of the PI while allowing authorized entities to access and process data. 
	The security standard applies to health plans, health care clearinghouses, and to any health care provider who transmits health information. 
	The security rule states that the PI must be confidential\footnote{Confidentiality of PI means that the information is not available or disclosed to unauthorized person\cite{hipaasecurity}.}, integral\footnote{Integrity of PI means that the information is not altered or destroyed in an unauthorized manner\cite{hipaasecurity}.}, and available\footnote{Availability of PI means that the information is accessible and usable on demand by an authorized person\cite{hipaasecurity}.}. 
	The PI holders must identify and protect against threats to the security and integrity of the information and protect from any impermissible uses or disclosures~\cite{hipaasecurity}. The security standard of HIPAA provides guidelines for the following safeguards along with non-compliance penalties: 
	\begin{enumerate}
		\item Administrative safeguards, including (i) security management process having risk analysis, risk management, sanction policy, and information system activity review, (ii) information access management having isolating health care clearinghouse functions, and access authorization, (iii) contingency plan having data backup plan, disaster recovery plan, and emergency mode operation plan.    
		\item Physical safeguards, including device and media controls with the implementation of disposal and media re-use provisions.
		\item Technical safeguards, including (i) access control having unique user identification, emergency access procedure, and encryption and decryption, (ii) audit controls having record and examine activity, and (iii) integrity control having mechanism to authenticate electronically protected health information.
		
	\end{enumerate}
	The HIPAA breach notification rule requires to provide notification following a breach of protected health information to the affected individuals. To strengthen the data protection requirements, there are other regulations along with HIPAA in the USA. These regulations include (a) Genetic Information Non-discrimination Act (GINA, \cite{GINA}), and (b) Health Information Technology for Economic and Clinical Health (HITECH, \cite{HITECH}). GINA was enacted in 2008, and it addresses the issues related to the discrimination based on genetic information, whereas HITECH was enacted in 2009, and it addresses the issues associated with electronic health records and health technologies. These regulations strengthen consumers' information rights on their data and prohibit disclosure of health information without their consent except for treatment, payment, or health care operations.

	
	\textbf{GDPR:}
	General Data Protection Regulation (GDPR)~\cite{gdpr} is the latest EU's data protection law. It has been in effect in EU since May 2018 to protect PI and harmonize data privacy laws across Europe and European Economic Area (EEA). It also regulates the PI data sharing outside EU and EEA. The GDPR applies to any organization that collects or processes PI of EU residents. It has the following six key principles~\cite{gdpr}: 
	\begin{enumerate}
		\item \emph{Lawfulness}, \emph{transparency} and \emph{fairness} while handling the PI. The organizations are obliged to inform the individual about the process of data handling transparently.
		
		\item The purpose of PI shall be \emph{specified}, \emph{explicit}, and \emph{legitimate}. Re-using the data for other purposes than the original one is restricted.
		
		\item The data storage and collection of PI shall be \emph{minimized} to that which is enough and relevant.
		
		\item The stored or collected data shall be \emph{accurate} and \emph{up to date} (by erasing or rectifying if the data is inaccurate).
		
		\item  The period of storing the PI shall be \emph{limited} to its necessity of the original purpose. It should be deleted once it is not necessary.
		
		\item PI shall be processed in a \emph{secure} manner, including protection against unauthorized or unlawful processing and accidental loss or damage. It is required that the data protection is ``\emph{by design}'' and ``\emph{by default}''. Privacy-by-design in data protection requires all safeguards necessary to ensure compliance with the regulation key principles since the first phases of relevant design and creation. On the other side, data protection by-default requires all steps to prevent unnecessary collection and processing of personal data other than needed for the purpose.
	\end{enumerate}
	GDPR empowers EU citizens by providing the right to access their PI, withdraw their consent at any time, ask to erase data, right to restrict processing, and right to be notified if their data is breached within 72 hours. Moreover, it also addresses the issues that can come due to the rise of ML algorithms in data processing. GDPR requires the explanations of the algorithmic outcomes before its implementation. Under GDPR, if the organizations do not comply with its regulations, then there is a provision of maximum penalties, including a fine that will be greater of \euro20 million or four percent of an organization's annual global revenue. EU has been working on Ethics for Artificial Intelligence, which includes fairness principle, transparent, intelligible, and responsible AI system, guaranteeing privacy by default, and by design~\cite{ethicsAI}.  
	
	
	\textbf{Australia Privacy Act:}
	In Australia, the Privacy Act 1988~\cite{privacyactaustralia1988} guides the privacy and security framework for PI. The Privacy Act applies to most Commonwealth government agencies (including tax office and department of human services), all private sector organizations that have an annual turnover of more than three million dollars, and some other organizations that meet particular criteria, for example, health service providers. Under the privacy act, the Australian Information Commissioner makes guidelines, known as the Australian Privacy Principles (APPs)~\cite{privacyactaustralia}. There are thirteen basic APPs, which are grouped into the following five parts. 
	\begin{enumerate}
		\item For the \emph{consideration of PI privacy}, there are two APPs. APP 1 outlines the requirements to manage PI openly and transparently. The organizations must have a clearly expressed and up to date privacy policy and complaints procedure. In addition, they must ensure their compliance with APPs. APP 2 states that the individuals must have an option of dealing anonymously or use pseudonym\footnote{A pseudonym is a name, term or descriptor that is different to an individual's actual name~\cite{privacyactaustralia}.} where possible.	
		\item For the \emph{collection of PI}, there are three APPs. APP 3 outlines that an organization can collect PI when it is reasonably necessary for, or directly related to, the organization's function or activities. The collection must be done lawfully and by fair means (with consent). APP 4 outlines the steps to take if the organizations receive unsolicited PI (collected without asking individuals). If the unsolicited PI can be collected under APP 3 and not in a Commonwealth record, then it must be destroyed or de-identified as soon as practicable. APP 5 outlines the information that must be provided to an individual when their data is being collected. This includes the organization's APP policy, detail about the organization such as contact details, purpose of the collection, complaint handling process, and potential overseas disclosure. 	
		\item To \emph{deal PI}, there are four APPs. APP 6 deals with the use and disclosure of PI. The PI can only be used or disclosed for the purpose it was collected or for a secondary use if an exception applies including consent for the secondary use, provide health services, and authorize by Australian law. APP 7 prohibits organizations from using or disclosing PI for direct marketing unless an exception applies. Direct marketing involves the use of PI to promote goods and services. If the organizations are allowed to use PI for direct marketing, then they must always allow individuals for an ``opting out'' option (not to receive direct marketing). Further, they must provide the source of PI to individuals upon request unless it is impracticable to do so. APP 8 introduces an accountability approach for cross-border disclosure. The organizations must ensure the overseas recipients follow APPs; otherwise, they may hold accountable for their recipient's breach. APP 9 restricts the adoption, use, and disclosure of government related identifiers\footnote{An identifier is a number, letter or symbol, or combination of any or all of those things, that is used to identify the individual~\cite{privacyactaustralia}.} by organizations.     
		\item For the \emph{integrity of PI}, there are two APPs. APP 10 requires organizations to ensure PI they collect, use, or disclose are accurate, up-to-date, complete, and relevant. APP 11 states that the organizations must take reasonable steps to protect PI they hold from misuse, interference, loss, unauthorized access, and modification, or disclosure (including hacking).    
		\item To \emph{access and correct PI}, there are two APPs. APP 12 requires PI holding organizations to provide individual access on request. It also sets out procedures for acceptance or rejection of the request. APP 13 states that the organizations must correct their data if it is wrong, or if an individual requests correction. This is required to ensure data accuracy, completeness, and relevancy. 
	\end{enumerate}
	Besides the Privacy Act, there is my health records act 2012~\cite{healthrecordact} in Australia. This act provides a legal framework for the management of my health record system, which provides an individual's key health information to the healthcare recipient. The My Health Records act follows the APPs for the collection, use, and disclosure of health information, including my health records. Besides, the Notifiable Data Breaches (NBD) scheme is commenced from February 2018 in Australia. Under this scheme, regulated entities (e.g., Australian Government agencies, business organizations with an annual turnover of \$3 million or more, health service providers and other organizations) require to notify affected individuals and the Australian Information Commissioner about the breaches that can harm one or more individuals. The overall summary of the requirements due to regulations is stated in Table~\ref{tab:legalrequirement}. 
	
	
	\begin{table}[th]
		\renewcommand{\arraystretch}{1.3}
		\caption{Summary of requirements due to regulations}
		\centering
		\resizebox{\textwidth}{!}{%
			\begin{tabular}{|c|l|l|}
				\hline
				Country                                                           & \multicolumn{1}{c|}{Requirements}                                                                                                                                                                                                                                                                                                                                                                                                                                                                                                                                                                                                                                                  & Overall requirements                                                                                                                                                                                                                                                                                                                                                                                               \\ \hline
				\begin{tabular}[c]{@{}c@{}}USA\\ (HIPAA~\cite{hipaaprivacy,hipaasecurity})\end{tabular}             & {\color[HTML]{000000} \begin{tabular}[c]{@{}l@{}}1) Consent (in written form) is required for any secondary use of health data.\\ 2) Health data must be kept confidential, integral, and available.\\ 3) Administrative safeguard, physical safeguard, and technical safeguard.\\ 4) Breach notification.\\ 5) HIPAA does not apply to de-identified health data.\end{tabular}}  
				
				& \multirow{-3}{*}{\begin{tabular}[c]{@{}l@{}}1) Proper consent\\ 2) Confidential\\ 3) Integrity and necessary updates\\ 4) Available\\ 5) Transparency and fairness\\ 6) Reasonable, minimum, and limited\\ 7) Secure and privacy-preserving\\ data processing\\ 8) Secure and compliance-friendly\\ data transfer\\ 9) Proper administrative, physical\\ and technical safeguard\\ 10) Breach notification\end{tabular}}
				\\ \cline{1-2}
				\begin{tabular}[c]{@{}c@{}}EU\\ (GDPR~\cite{gdpr})\end{tabular}               & {\color[HTML]{000000} \begin{tabular}[c]{@{}l@{}}1) Lawfulness, transparency, and fairness.\\ 2) Minimum and limited data storage and collection, integrity check, and\\  up to date.\\ 3) The secure processing, data security including two-factor authentication\\  and end-to-end encryption, data protection by design and by default.\\ 4) Secure and compliance-friendly data transfer.\\ 5) Freely given, specific, informed and unambiguous consent, and data\\  subject rights to their data.\\ 6) GDPR applies to pseudonymized data if the data subject can be\\  identified by linking other additional available information.\\ 7) Breach notification\end{tabular}} &                                                                                                                                                                                                                                                                                                                                                     \\ \cline{1-2}
				\begin{tabular}[c]{@{}c@{}}Australia\\ (Privacy Act~\cite{privacyactaustralia1988})\end{tabular} & {\color[HTML]{000000} \begin{tabular}[c]{@{}l@{}}1) Open and transparent management of personal information along with\\  anonymity and pseudonymity if not exempted.\\ 2) Reasonable and lawful collection of information only with consent and\\  notification.\\ 3) Prohibit disclosing PI, including marketing and cross-border, without\\  proper consent, and other tasks than the original purpose.\\ 4) Security and integrity.\\ 5) Access control and data correction.\\ 6) Breach notification.\end{tabular}}   &                                                                                                                                                        
				\\ \hline
			\end{tabular}
		} \label{tab:legalrequirement}
	\end{table}
	
	
	For all countries besides their laws, there are binding international laws such as the universal declaration of human rights and the European convention on human rights~\cite{wholegalframwork}. These laws stress on the privacy of the individuals (e.g., PI), and their requirements are covered above.   
	
	
	\subsection{Requirements due to ethics}
	
	Ethical guidelines and framework are to ensure the responsibility of collection and analysis of PH data for any purpose. These enable us to decide the use of appropriate technology for social good. Recently, health (big) data analytic, which relies heavily on AI, has brought the ethical concerns of PH data more than ever on privacy, control, and data ownership. More precisely, ethical issues include the possibility of re-identification of users by linking, merging, data-mining, and re-using datasets in volume. 
	
	The study performed by an Italian company named Evodevo srl, with the support of the European Economic and Social Committee, explores the ethical dimension of Big data, which also includes PH data. It states the following ethical issues~\cite{EESC}:
	
	\begin{enumerate}
		\item Awareness: Lack of awareness can lead to unethical use of health data.
		\item Control: To provide true ownership of the user's health data, users need to have control, including removal, of their private data provided to the service provider, including the data provided to other parties by the service provider.
		\item Trust: Trust is required for the user's acceptance to provide personal data for a service (e.g., health advice).
		\item Ownership: It is necessary to clearly state the ownership related to the data after processing the original user's data.
		\item Surveillance and security: Unnecessary surveillance to limit the citizen's liberty is unethical.
		\item Digital identity: The online profile of individuals due to his/her online activities can be used for discrimination.
		\item Tailored reality: Focused and targeted service based on personalized information, including advice and advertisement, limits the user's exposure.
		\item De-Anonymization: There are concerns related to linking the information from two or more sources to infer more information from the de-anonymized data.
		\item Digital divide: Digital divide refer to an inability to use the new technologies (e.g., by senior citizens) for the services delivered through the new technologies.
		\item Privacy: Privacy is required to prevent the use of health data without consent.
	\end{enumerate}

	The relevant requirements based on the issues mentioned above on PH data related to privacy, security, and trust are presented in the following:
	
	\begin{enumerate}
		\item \emph{Awareness, control}, and \emph{ownership}: It is required to practice informed use of data, provide user control over his/her data not only to the primary custodian, but also secondary custodians (which got the data from the primary custodian), and clear ownership criteria for the evolved data after processing the original user's data. 
		\item \emph{Trust}: Ethical guarantees for the usage of data is necessary to gain trust from users.
		\item \emph{Privacy}: Privacy aspects are essential to preventing data usage without proper consent and approval, specifically, for secondary data usage.
		\item Limiting the \emph{information linkage}: It is required to preventing linkage of the data of one source to data from other sources to infer more information on the data subject other than the original intention. Thus, while releasing the health data to the public, proper consideration must be carried out such that there will be no or less possibility of extraction of sensitive information from the linkage.
	\end{enumerate}

	National statement on ethical conduct in human research, Australia, states the following requirements on human data use in research~\cite{NSEC}: 
	\begin{enumerate}
		\item \emph{Ethics approval} from the designated ethics committee: Ethics approval is the first step before conducting human research activities, including collection, store, and analysis of human-related data (such as PH data).
		
		\item \emph{Consent}: Consent should be taken voluntarily from the participants after providing adequate information about the proposed research and implications of the participation. Renegotiation of consent is required if the original terms change over time. Participants have opportunities to decline or withdraw consent. There are three types of consent in research, namely specific, which is limited to a particular project, extended and unspecified, which are given for the use of data in future research projects.
		
		\item Address the following ethical issues related to the collection, use, and management of human data and information:
		\begin{itemize}
			\item \emph{Identifiability} of information: It is required to de-anonymized the human data based on the requirements of the research, and proper care must be taken to reduce the likelihood of re-identification of individuals during collection, analysis, and storage of data.
			\item \emph{Data management}: Proper access control and usage (e.g., analysis and re-use) are required if multiple researchers are collaborating on the same human data repository. Besides, other required measures for data management include physical, network, and system security, confidentiality agreement, and safe disposal.
			\item \emph{Secondary use} of data or information: It is required to re-obtain the consent wherever applied. If unpractical to do so, then the usage must be ethically justified.   
			\item \emph{Data sharing}: It is required to follow the data management plan and ethical norms (including re-consenting if required and confidentiality agreement) while sharing data with other researchers.   
			\item Dissemination of \emph{project outputs} and outcomes: The dissemination of the inferences from the human data, including the outputs, need to be aligned with the ethical principles (e.g., the privacy of the participant).     
		\end{itemize}
		\item \emph{Risk analysis and management}: At various stages in research, there are risks associated with the privacy of human data and information. These risks need to be analyzed and managed properly at that stage.    
		
	\end{enumerate}

	Based on the American Medical Association code of medical ethics~\cite{AMA}, we find the following requirements for the health data: 
	\begin{enumerate}
		\item  \emph{Privacy}: A comprehensive privacy, including physical privacy, information privacy, decision privacy, and associational privacy, is required.
		\item \emph{Confidentiality}: For confidentiality, it is required to restrict disclosure of the data to third-party. If disclosure is needed for the benefits of the data subject, then only the minimum necessary information should be disseminated, considering re-obtaining consent if applicable. Any third party can have access only to de-identified information. Besides, the duty of confidentiality extends beyond the death of the data subject.
		\item \emph{Medical records management}: It is necessary to safeguarding and monitoring the confidentiality of the patient's personal information, proper access control mechanism to the data, its storage time limitation, and availability of the record if requested by a patient, or in point of care by a physician.
		\item \emph{Breach notification}: The patients must notify about the data breach if occurred. 
	\end{enumerate}
	
	In various other ethical guidelines, including the world medical association declaration of Helsinki - Ethical Principles for Medical Research involving Human Subjects~\cite{WMAethics}, and a document by world health organizatin~\cite{wholegalframwork}, primarily stress on the (i) privacy and confidentiality of personal information, and (ii) informed consent.

	\begin{table}[]
		\renewcommand{\arraystretch}{1.1}
		\caption{Summary of requirements due to ethical guidelines}
		\resizebox{\textwidth}{!}{%
			\begin{tabular}{|l|l|l|}
				\hline
				\multicolumn{1}{|c|}{References}   &\multicolumn{1}{c|}{Requirements}                                                                                                                                                                                                                                                                                                                                                                                                   & \multicolumn{1}{c|}{Overall requirements}                                                                                                                                                                                                                                                                                                                                                                                                                            \\ \hline
				
				\begin{tabular}[c]{@{}l@{}}Evodevo srl~\cite{EESC}\\ (supported by\\ EU Economic and Social Committee)\end{tabular}            & \begin{tabular}[c]{@{}l@{}}1) Awareness, Control and Ownership\\ 2) Trust (platform)\\ 3) Privacy\\ 4) Measures to limit the information linkage\end{tabular}                                                                                                                                                                                                                                                                                  & \multirow{4}{*}{\begin{tabular}[c]{@{}l@{}}1) Awareness \\ 2) Control\\ 3) Ownership\\ 4) Trust (platform)\\ 5) Privacy, security, and confidentiality\\ 6) Ethics approval in research projects\\ 7) Informed consent with the flexibility\\ to opt-out and transparency\\ 8) Proper de-anonymization to limit the \\ information linkage from various sources\\ 9) Proper data-sharing management\\ 10) Risk analysis and management \\ 11) Breach notification\end{tabular}} \\ \cline{1-2}
				
				\begin{tabular}[c]{@{}l@{}}National statement on ethical conduct\\ in human research, Australia~\cite{NSEC}\end{tabular}          & \begin{tabular}[c]{@{}l@{}}1) Ethics approval\\ 2) Informed consent with an opportunity to decline or withdraw it\\ 3)  Measures to limit the re-identification of  individual\\ 4) Data management for security, confidentiality, and privacy\\ 5) Ethically justified or consented use of data for secondary purposes\\ 6) Ethical data-sharing  management, including the outputs\\ 7) Risk analysis and management\end{tabular} &                                                                                                                                                                                                                                                                                                                                                                                                                                                                      \\ \cline{1-2}
				\begin{tabular}[c]{@{}l@{}}American Medical Association code\\ of medical ethics~\cite{AMA}\end{tabular}                                                                     & \begin{tabular}[c]{@{}l@{}}1) Privacy\\ 2) Confidentiality\\ 3) Medical records management, security, and privacy\\ 4) Breach notification\end{tabular}                                                                                                                                                                                                                                                                             & 
				\\ \cline{1-2}
				\begin{tabular}[c]{@{}l@{}}World medical association declaration\\ of Helsinki~\cite{WMAethics}, and\\ World health Organization~\cite{wholegalframwork}\end{tabular} & \begin{tabular}[c]{@{}l@{}}1) Privacy and confidentiality \\ 2) Informed consent\end{tabular}                                                                       &
				\\ \hline
			\end{tabular}%
		}
		\label{tab:ethicalrequirement}
	\end{table}

	In overall, Table~\ref{tab:legalrequirement} and~\ref{tab:ethicalrequirement} provides the requirements for the legal and ethical compliances, respectively.

	\subsection{Requirements due to health domain}
	So far, we have discussed the requirements of the security, privacy, and trust from the data usage, collection, and storage perspective. Now, we present the domain-specific need that is not covered above. As the data are crucial for medical decision making, the data must be accurate, complete, and precise to avoid wrong decisions. An incorrect medical decision can harm the patients up to their death; thus, proper care must be taken. This results in the requirement of the PH data trustworthiness. In this paper, the requirements due to the issues related to the trustworthiness of health devices~\cite{devicetrust}, which is also an essential issue in the medical domain, are out of scope. We explore only up to data trustworthiness.
	
	The PH data trust is essential for the precision health that envisions to provide health care based on evidence derived from the data. We found a lower number of documents explicitly stating the requirements for (PH) data trust. Based on our readings, including data trust~\cite{datatrust}, health data quality~\cite{whodataquality,AHIMA}, and FAIR principle~\cite{FAIR}, the requirements are the following:
	\begin{enumerate}
		\item \emph{Standard format}: Health data stored in a proper format across different organizations ensure ease of data processing, e.g., data matching, and data integration. Besides, the metadata and data should be easily human and machine searchable. There is a need for machine-readable metadata (e.g., proper indexing) maintained along with the data if the data search is automated. 
		\item \emph{Simple, clear} and \emph{complete}: Present data in a way such that further analysis and inference is consistent if it is processed at a different time or by different organizations.  
		\item \emph{Accurate, timely,} and \emph{transparent}: The health data should be accurate (e.g., correct data entry and updates). It should have adequate additional information to verify its credibility, including source information, data entry, and collection method. Transparency is also required to check the correctness of the data. The data should be recorded and processed on time to avoid possible errors and incompleteness.     
	\end{enumerate}

	\section{Major challenges in precision health data security, privacy and trust}
	\label{sec:challenges}
	The following are the major challenges for PH data security, privacy and trust: (1) Health data security and privacy whilst computing, (2) consent management, (3) PH data trustworthiness, and (4) legal and ethical compliance.

	\subsection{Health data security and privacy whilst computing:}
	Security and privacy of data-at-rest are ensured by well-defined encryption methods such as Advanced Encryption Standard (AES)~\cite{aes}\footnote{AES is quantum-safe as the AES-encrypted data can resist quantum attacks by increasing its key size. A cryptographic protocol is said to be Quantum-safe if it is well examined under all known quantum algorithm.}, Rivest-Shamir-Adleman (RSA)~\cite{RSA}, Elliptic Curve Diffie-Hellman (ECDH)~\cite{ellipticcurve}. 
	Besides, query operations over the encrypted data (e.g., encrypted data-at-rest) without decryption is possible due to searchable encryption~\cite{searchableencrypt}.
	Data-in-transit is protected by secure protocols such as Transport Layer Security (TLS)~\cite{TLS} and File Transfer Protocol Secure (FTPS)~\cite{FTPS}. On the other hand, protecting data security and privacy for data-in-use is a difficult task as it is associated with data computation, and whilst computation, the data usually requires decryption revealing itself to the computing platform. Moreover, the computing platform may not be a trusted platform. There are various evolving techniques to handle this issue, such as trusted platform, homomorphic encryption (computation over encrypted data), and multi-party computation. However, either they require a trusted vendor or they do not have product-ready protocols. These approaches are subjected to more research and development for its wider use. 
	
	Despite these challenges, there has been continuous progress in confidential computing, and related products and services are available in the market. Usually, these services are based on trusted platforms and memory encryption to isolate the data whilst computation, and provided by big technology companies, including Microsoft (Azure confidential computing) and Google (Google cloud confidential computing). However, recent studies show that these trusted platforms can be vulnerable to attacks, such as side-channel and timing. Refer to Section~\ref{sppreservingtech} for details in security and privacy for data-in-use.

	\subsection{Consent management:}
	Consent is mandatory for health data handling, including collection, analysis, and storage. It is an important tool to protect individual privacy, confidentiality, and autonomy. This is governed both by ethical guidelines and legislation. There are three types of consent, namely \emph{explicit}, \emph{implicit} and \emph{opt-out} consent. In explicit consent, the purpose of collecting personal information, its use, handling, and disclosure of the information are presented with an option to agree or disagree. This type of consent is required for all aspects of clinical trials, including the retention of medical records. This is also called \emph{opt-in} consent. It is used whilst handling the information. In implicit consent, consent is deemed in favor of both the data subject and collector. Most of the cases, this consent is obvious at the time of collection (e.g., a doctor taking blood samples of his patient for lab tests). In an opt-out consent, the participants are informed about the purpose of consent with an option to decline it. If it is not declined, then the consent is considered to be provided. A consent management solution for enterprises has been proposed by researchers at IBM~\cite{ConsentmanagmentIBM}. This solution provides tools for modeling consent, a repository for storing it, and a data access management component to enforce consent and log the enforcement decisions.

	The main problem related to consent arises whilst data sharing and data linkage. This is usually required in the data pre-processing phase of health data analytics, where data come from various sources (e.g., hospital, insurance company, and social media). There are two approaches for consent, namely static consent and dynamic consent. In static consent, the consent must be taken for all future usage of data at the time of data collection, and it is usually paper-based. It cannot address the issues that come with the change in environment and requirements with time, such as reusing the data for a different health project other than originally consented. In this regard, dynamic consent~\cite{dynamic_consent,dynamicconsent} is advantageous. Dynamic consent is an informed and personalized consent, where two-way communication is interfaced between the data subject and data custodian, and the subject can update and provide different kinds of consent. In addition, the subject can control their health data usage over time and revoke consent through the interface. Besides, the consent is traveled with the corresponding data when it is shared with other parties, and also, the participant can get the research results. However, dynamic consent has challenges, including higher implementation cost,  consent revocation, and data deletion guarantee, and need to have patients with sufficient digital knowledge and time. Overall, how to automate the consent and manage it efficiently in the interest of legislation, patient's autonomy, cost, and data analytics is still an open problem. 
	
	The health data analytics require more health data, that means more participants and their consent, for better health care quality. Thus, it is equally important to explore the approaches that increase consent approvals. One possible way to do so is through trust (trustworthy system), as a study suggests that trust and privacy concerns are inversely proportional to each other~\cite{trust}.

	\subsection{PH data trustworthiness}
	As health data is complex and diverse, checking and maintaining the trustworthiness of health data is a considerable challenge. In addition, the increasing size of health data (e.g., big data), distributed storage of health data (e.g., hospitals and pharmaceuticals) at different places, and a massive number of data sources (e.g., medical internet of things) add additional difficulties and complexities for checking the trustworthiness. Credible sources such as government agencies and reputed organizations are trustworthy data sources. They follow health data governance policies so that one can inspect their data via metadata and associated information. However, in the advent of the internet of medical things (IoMT), e.g., smartwatch monitoring heart rate, it is difficult to manage and maintain the reliability of health data, where data can be extracted from a faulty or improperly configured IoMT device.

	\subsection{Legal and ethical compliance:}
	Legal and ethical compliance is necessary while handling PH data. Otherwise, there may be a trust problem or hefty fine (e.g., \euro20 million or four percent of an organization's annual global revenue as stated by GDPR in EU) for the breach. To understand the privacy risks when conducting data processing (e.g., data analytic) and possible ways to reduce them for compliance, we present a summary of the guidelines presented in ``Guide to data analytics and the Australian privacy principles''~\cite{dataanalyticsAPP} as an example. Refer to Table~\ref{APPguide} for details.         
	
	\begin{table}[!ht]
		\renewcommand{\arraystretch}{1.3}
		\scriptsize
		\caption{Privacy risk factors and possible risk reducing steps}
		\label{APPguide}
		\centering
		\begin{tabular}{|p{7cm}|p{8.5cm}|}
			
			\hline
			\hskip60pt Privacy risks &  \hskip50pt Possible risk reducing steps\\
			\hline
			Data may contain personal information, and it is subjected to the Privacy Act.	
			& Proper de-identification of data. \\
			\hline
			No proper de-identification.
			& Risk assessment to consider the likelihood of re-identification, and implement risk mitigation techniques.\\
			\hline
			Privacy impact assessment (PIA)~\cite{PIA} is challenging for big data.
			& PIA needs to be carried out. \\
			\hline
			Using `all the data' for `unknown purposes'.
			& Limit the collection and use of personal information to a reasonably necessary level to perform legitimate functions. \\
			\hline
			New personal information creation during analytics.
			& If not legally collected, then needs to be de-identified and destroyed. \\
			\hline
			Information collected by third party included in analytics.
			& Follow the consent provided for secondary use of those information. \\
			\hline
			People do not read privacy notices.
			& Customize the notices to make them easy, dynamic and user friendly. \\
			\hline
			Secondary use and disclosures of personal information are common in data analytics.
			& Check compatibility with the original purpose of collection or rely on exceptions. Send privacy notices to inform individuals about the particular use or disclosure. \\
			\hline
			Impracticable to obtain individuals’ consent.
			& Follow the law and guidelines (e.g., Australian Government National health and Medical Research Council's guidelines~\cite{nhmrc} whilst handling personal health information).\\
			\hline
			Personal information disclosure to an overseas recipient.
			& Adopt extra diligence and follow law before disclosure.\\
			\hline
			Algorithmic biases in its decisions which are discriminatory, erroneous and unjustified.
			& Ensure correctness of models and methods. \\
			\hline
			Information collected from third party may not be accurate, complete and up-to-date.
			& Take rigorous steps to ensure the data accuracy, correctness and updates.\\
			\hline
			Hacking risks.
			& Take proper security and prevention measures (e.g., data encryption, controlled access, and network security).\\
			\hline
		\end{tabular}
		\vskip-3pt
	\end{table}

	A compliance check is a difficult task, especially when data is collected from various sources, including third parties, and it is collected in a huge amount (usual case in big PH data). In addition, the law can be vague, and ethics are highly conceptual and abstract. It is unclear how to effectively and automatically check the compliance for data-in-use cases. However, by vigilant inspections, using proper platforms (e.g., privacy-by-design and privacy-by-default), auditing (e.g., privacy impact assessment~\cite{PIA}), using compliance analytics\footnote{Compliance analytic calculates and prioritizes risk factors, and identifying highest risk transactions. These insights are used to manage the compliance risk.}, and strictly following a compliance checklist in each data processing step starting from data collection to final output predictions, one can self-regulate the check and reduce the possible risks.

	So far, we have explored the requirements due to regulations, ethics, and data trustworthiness for the PH data. We have identified that data security and privacy whilst computing is one of the main challenges. By considering its high impact in precision health, our survey is limited up to it in the remainder of this work; other major challenges are excluded. In this regard, we present the best available security and privacy-preserving techniques, including ML/AI techniques. These techniques ensure ethical and regulatory requirements while handling and using PH data in the healthcare domain. 
	\color{black}
	\section{Techniques for PH data privacy and security} \label{sppreservingtech}
	Security and privacy of PH data or information is a mandatory requirement for health databases, including personal information (PI), worldwide due to legal provisions, financial reasons, and trust. In this section, we briefly discuss security and privacy-preserving techniques that are relevant to PH data. As one technology alone cannot provide a complete security and privacy solution, combinations of more than one are required.
	
	\subsection{Security and privacy for data-at-rest and data-in-transit} \label{sec:datasecurityandprivacy1}
	\subsubsection{Data security:}
	There are four primary techniques for data security. These techniques are (1) cryptographic security, (2) blockchain-based security, (3) access control and security analysis, and (4) network security.   
	\paragraph{Cryptographic security:}
	Cryptography~\cite{cryptographybook} is an essential technique for data security against interception, tampering, and unauthorized reading. It deals with various data security aspects including \emph{authentication} (checking and confirming the identity), \emph{integrity} (ensuring only an authorized user makes modifications to the data), \emph{confidentiality} (allowing only authorized recipients access the data), and \emph{non-repudiation} (preventing the denial of earlier commitments or actions)~\cite{cryptographybook}. Data encryption plays a vital role in protecting sensitive information. However, its implementation is not extensive. According to Gemalto, in the first half of 2018, only 2.2\% of the total data-breach incidents (worldwide) had data in encrypted form~\cite{breachlevelindex} (useless for the breacher). In the same report, health data breach accounts for 27\% with the highest among all breach incidents by industry.  
	
	Encryption can be both software-based (e.g., Microsoft Windows BitLocker~\cite{bitlocker}, VeraCrypt~\cite{veracrypt}) and hardware-based (e.g., Seagate secure self-encrypting hard drives~\cite{seagateselfencrypting}). As the encryption and decryption are carried on by dedicated hardware components (not the main processor) in hardware-based encryption, it is faster than its software counterpart. Moreover, the encryption keys are stored locally (inside disk) in hardware encryption, which makes it more secure than software encryption, where keys can present in random access memory (RAM) locations whilst processing. Attacks such as cold boot attack can read keys present in RAM~\cite{coldbootattack}. For disk storage devices, there are various storage encryption technologies, including full disk encryption, virtual disk encryption, volume encryption, and file/folder encryption, which provide different levels of securities~\cite{storageencryption}. In a computing environment, disk encryption is not sufficient as information can be leaked from a processor or memory if it is not encrypted there. Thus memory encryption~\cite{memoryencryption} and cryptoprocessors~\cite{cryptoprocessor} are implemented along with disk encryption.        
	
	\paragraph{Blockchain-based security:}
	Blockchain is a distributed public ledger that maintains a sequentially growing list of transactions or data in a chain of blocks. The information inside the blocks are immutable (no single party can delete it) and time-stamped. Blockchain enables data sharing without trusting the compute-nodes of a network, and a central node does not control it. Blockchain enables the security of networks and systems via data integrity. For example, Keyless Signature Infrastructure (KSI) blockchain~\cite{ksi} enables secure, scalable, digital, signature-based authentication for electronic data, machines, and PI. Also, KSI blockchain is quantum-safe. Estonia health care system~\cite{ksiwebsite} and Personal Care Record Platform, called MyPCR~\cite{MyPCR}, use KSI blockchain to ensure data integrity and security in their system. Blockchain can be used for patient-driven healthcare interoperability. It can facilitate various aspects of interoperability, including digital access rules management, data aggregation, data availability and liquidity, patient identity, and immutability, though with some limitations, including handling the big PH data, privacy, security, and incentives considerations~\cite{gordon}.       
	To maintain the end-to-end confidentiality of genomic data queries, blockchain is used together with homomorphic encryption and secure multi-party computation~\cite{dennis}.
	Despite the ability to improve data security due to the data encryption on blockchain, there are possibilities of PI leakage from a public blockchain due to attacks, including linkage attacks~\cite{agbo}. The scalability of blockchain, specifically, for the big PH data, is another primary concern.

	\paragraph{Access control and security analysis:}
	It is essential to secure physical devices (including desktop computers, laptops, and tablets) and infrastructures (including healthcare facilities, healthcare cloud servers, and data centers) that are holding sensitive private data. According to Verizon's 2018 data breach investigation report~\cite{verizon}, 11\% of the total data breaches involved physical actions, including theft of physical devices and paper documents. If an intruder gets access to those devices or infrastructures (premises), then he can obtain sensitive private data (i.e., data theft) and can damage the infrastructure or data (i.e., data loss). Other physical security risks include natural disasters, including fire, earthquake, and flood. A proper secure data backup system is necessary to recover the data when these risks occurred or system failure. Besides, access control mechanisms, which is a conventional approach to data security, regulate the users and their access to sensitive data. It performs identification authentication and authorization of users. Multi-factor authentication using passwords, bio-metric scans, cryptographic tokens, and RFID cards are standard mechanisms for access control. Besides, real-time security analytics device/software such as Intrusion Detection System (IDS)~\cite{intrusiondetection} and Intrusion Prevention System (IPS)~\cite{intrusionprevention} are essential security measures.

	\paragraph{Network security:}
	Network security maintains the security and privacy of data-in-transit. It is maintained through security protocols and standards such as Secure Socket Layer (SSL), Transport Layer Security (TLS), Secure HTTP, secure IP (IPsec), and Secure Shell (SSH). TLS and SSL provide transport-level security, IPsec provides network-level protection, and Secure HTTP offers secure communication between a HTTP client and a server~\cite{networksecurity}. Protocols such as wired equivalent privacy (WEP) and Wi-Fi protected Access (WPA) protects wireless networks. Besides, an untrusted network such as the internet (a public network) consisted of security threats including computer virus, Trojan horse, adware, spyware, worm, and rootkit. A firewall, which is a network security system, and IPS (e.g., antivirus software) enable security in a local network that is connected to the untrusted network by monitoring and controlling all incoming and outgoing network traffic of the local network.

	\subsubsection{Data privacy:}
	There are mainly three risks to data privacy, namely singling out, linkability, and inference~\cite{EUanonymisation}. Singling out refers to the identifying individual/attribute/value in a dataset by isolating the records. In contrast, linkability refers to identifying an individual/attribute/value in a dataset by linking two or more other files related to the same individual/attribute/values. On the other side, inference refers to the possibility to identify the individual/attribute/values from the different individuals/attributes/values with a significant probability. 
	The two primary techniques for data privacy are (1) anonymization and (2) pseudonymization.   
	\paragraph{Anonymization:}
	Anonymization includes randomization and generalization~\cite{EUanonymisation}. Randomization techniques modify the integrity of the data to avoid the active link between the data and the individual. On the other hand, the generalization technique generalizes or dilute the attributes of data subjects by changing the respective scale or order of magnitude. For example, writing region instead of the street, and a range of years rather than a specific year. Randomization is used against inference attacks, but not effective against singling out and link attacks. In contrast, generalization is effective against singling out but requires quantitative approaches to prevent linkability and inference. Randomization techniques~\cite{EUanonymisation} include noise addition (retain the overall distribution but hide individuals), permutation (shuffling the values of attributes in a table such that some of them are intentionally linked to different data subjects), and differential privacy (robust but there is a trade-off between the usability and anonymization, see Section~\ref{sec:diffprivacy}). Generalization techniques~\cite{EUanonymisation} include aggregation, K-anonymity, and L-diversity/T-closeness. Aggregation and K-anonymity protect against singling out by grouping them with, at least, K other individuals. On the other side, L-diversity is the extension of K-anonymity such that, in each equivalence class, every attribute has at least L different values to avoid inference attacks. And, T-closeness is the improved L-diversity such that equivalent classes resembling the initial distribution of attributes in the table are created to keep the data as close to the original one. Despite various techniques in anonymization, it is shown not sufficient for the privacy guarantee in a recent work~\cite{natureanonymization}.   
	
	\paragraph{Pseudonymization:}
	Pseudonymisation~\cite{EUanonymisation} replaces one attribute in the dataset by another to reduce the linkability between the original identity of a data subject and the dataset. The techniques for pseudonymization include 
	\begin{itemize}
		\item encryption with a secret key,
		\item hash function (a function that returns a fixed-size output from an input of any size and cannot be reversed), 
		\item keyed-hash function with stored key (a particular hash function that uses a secret key as an additional input),
		\item deterministic encryption (a keyed-hash function with deletion of the key), and 
		\item tokenization and masking (it replaces a part of data by a random or semi-random data, called token, which retains the format and data type of the replaced part of the data). For example, a dynamic data masking (MAGEN) by IBM~\cite{MAGENbyIBM} implements data masking to allow data sharing whilst safeguarding sensitive business data.   
	\end{itemize}

	
	\subsection{Data security and privacy for data-in-use} \label{sec:datasecurityandprivacy2}
	In this section, we introduce some important evolving data security and privacy-preserving techniques that are relevant to precision health, specifically for data-in-use cases. We discuss these techniques and their implementation in the healthcare domain. The summary of these techniques is presented in Table~\ref{sumtable}.   
	
	\subsubsection{Trusted Execution Environment:} \label{trustedexecutionenv}
	Trusted Execution Environment (TEE) provides secure storage and isolation of sensitive computations from other processes, including operating systems, BIOS, and hypervisor. Moreover, it reduces the attack surface by isolation and cryptography, and thus increases the security of the processes running in TEE. TEE uses a hardware module or software module or both modules for the confidentiality and integrity of data and application code. Moreover, it has a mechanism for remote attestation that provides proof of trustworthiness to the users. TEE considers the threat model that includes all software attacks and physical attacks on the main memory and its non-volatile memory. There have been several works from industry and academia on providing TEE. These works include ARM TrustZone~\cite{trustzone}, Intel SGX~\cite{intelsgx}, Trusted Platform Module~\cite{tpm}, Intel TXT~\cite{txt}, AMD Security Technology~\cite{sev}, Sanctum~\cite{sanctum}, and Keystone Enclave~\cite{keystone}. For more insights into TEEs' environment and developments, we discuss some notable TEEs in the following. 
	\begin{itemize}
		\item ARM TrustZone~\cite{trustzone}:
		ARM TrustZone is a hardware level technology that enables the ARM processor system into two hardware-isolated zones, namely trusted zone and non-trusted zone. Both zones have their operating system and data. System modules like drivers and applications do not have direct access to the trusted zone. It is separated from the normal world operations, and thus from the attacks exploiting the normal resources. The trusted zone handles the sensitive operations and data that need to be secured. The secure context switching between the two zones is managed by special software called \emph{secure monitor} in the case of Cortex-A processors, and a set of mechanisms (precisely three instructions, namely secure gateway, branch with exchange to non-secure state, and branch with link and exchange to non-secure state) implemented into the core logic in the case of Cortex-M processors. Several academic research works and commercial products have used ARM TrustZone based TEEs~\cite{trustzonesurvey}. Though the TrustZone is a key enabler for the development of trustworthy systems, it is vulnerable to various attacks, including those exploiting bugs in the TEE kernel, hardware exceptions, caches, and power management modules~\cite{trustzonesurvey}.

		\item Intel SGX~\cite{intelsgx}:
		Intel Software Guard Extensions (SGX) are instruction-set architecture extensions that provide a trusted computing environment by leveraging trusted hardware. It uses secure containers, called \emph{enclave}, for the protection and isolation of its contents (code and data) from other processes and other enclaves. The memory is encrypted with a key that is unique to each enclave. The enclaves are trusted components. Intel SGX provides a software attestation method that allows a remote client to authenticate the program executing inside an enclave. The implementation of Intel SGX in the real world for the development of various applications is made possible due to the availability of software development kits such as Intel Platform Developers Kit~\cite{intelsdk}, Fortanix Enclave Development Platform~\cite{fortanix} and Open Enclave SDK~\cite{opensdk}, and cryptographic library such as Intel SGX SSL library~\cite{intelssl} dedicated for SGX. Unlike ARM TrustZone, Intel SGX has data sealing features and memory protection from physical attacks such as bus probing. However, one needs to trust the vendor fully. Besides, Intel SGX is vulnerable to side-channel attacks, including those exploiting page tables, caches, translation lookaside buffer, and DRAMs used by enclave programs~\cite{intelsidechannelattack}. These attacks can be mitigated by using Compiler/SDK techniques and Microcode patch. 
		
		\item Keystone Enclave~\cite{keystone}:
		Keystone Enclave is an open-source enclave for RISC-V processors. RSIC-V is an open-source hardware instruction set architecture (ISA). Keystone Enclave uses hardware capabilities in RISC-V to design a secure enclave. In contrast to the commercial and proprietary TEE environment (e.g., Intel SGX), open-source TEE environment provides transparency and inside details. This results in an experiment and research openly by academia and industry to address challenges in enclaves, including hardware vulnerabilities and side-channel attacks. Keystone security monitor, a special module running in machine mode (trusted mode), manages enclaves, Physical Memory Protection (PMP) entries, multi-core PMP synchronizations, and remote attestation. Memory and its buses are encrypted for the defense against physical attacks. Keystone Enclave has strong memory isolation enabled by using separate virtual memory management (other than that of Operating System) and ISA-enforced memory access management. It is a relatively new execution environment, and researchers are still working in its improvement and building software stacks, including toolchain and edge compilers.

	\end{itemize}
	It is still possible to emerge new attacks and defense mechanisms for TEEs. However, in all cases, the attackers should have privileged access or specific condition, which is not common in general. 
	
	\paragraph{TEEs and healthcare:}
	Privacy and security of PI are of prime concern in the healthcare domain, and it is the most common use case of TEEs. Intel SGX is used to increase the trust and security of health data exchange in a Horizon 2020 project (a European Union Research and Innovation program) named KONFIDO~\cite{usecaseintelsgx}. In KONFIDO, decryption, transformations, and encryption of patient summaries were carried out in the TEE provided by SGX. 
	In another work, a privacy-preserving international collaboration framework for analyzing rare disease genetic data is introduced~\cite{usecaseintelsgx3}. This work leverages Intel SGX for trustworthy computations over distributed and encrypted genomics data. ARM TrustZone technology has been implemented to secure the medical internet of thing devices~\cite{arm}.  

	\subsubsection{Homomorphic Encryption:} \label{homoenc}
	Homomorphic Encryption (HE) allows the computations (arbitrary functions) over encrypted data without decryption. The computing environment would not be able to know the data and results, which both remain encrypted. Thus, HE enables secure computation on an untrusted computing platform. Depending upon the number of allowed operations on the encrypted data, there are three types of HE~\cite{homomorphicsurvey}, which are as follows:
	\begin{enumerate}
		\item Partially homomorphic encryption: Partially homomorphic encryption (PHE) allows only one type of operation on the encrypted data for an unlimited number of times. It supports either only addition or multiplication. Some examples of PHE schemes are RSA~\cite{RSA}, GM~\cite{GM}, and KTX~\cite{KTX}.
		\item Somewhat homomorphic encryption: Somewhat homomorphic encryption (SWHE) allows more than one type of operation on the encrypted data but only up to a certain complexity and for a limited number of times. It supports both addition and multiplication, but the number of HE operations is limited because of the size of the ciphertext increase, and noise gets accumulated with each HE operation. Some examples of SWHE are Yao's Garbled circuit~\cite{yao}, SYY~\cite{syy} on NC1 circuits, and IP on branching program~\cite{IP}.   
		\item Fully homomorphic encryption: Fully homomorphic encryption (FHE) allows any operations on encrypted data for an unlimited number of times. Gentry~\cite{FHEGentry} first proposed a general framework for FHE. His scheme was based on ideal lattices. Further improvements in FHE schemes have been observed in several following works~\cite{homomorphicsurvey}.   
	\end{enumerate} 
	
	HE, especially FHE, plays a vital role in the privacy and security of PI. Its real-world implementation is challenging in general due to high computational requirements and overhead. Optimization has been done based on its use cases~\cite{fasterFHE,fasterFHE2}, but it is still insufficient for a general case. The implementation of a fully functional FHE, whilst large data of different structures are input from multiple sources with different encrypting key, is an open problem. This type of environment is prevailing in precision health platform. Homomorphic encryption does not provide verifiable computing, so it should use other mechanisms for the purpose. For collaborative computations, HE can suffer from a collusion attack because all parties share the same public key, and the dishonest party can collude with the server~\cite{privacypreservingcomp}.   
	
	\paragraph{HE and healthcare:}
	As a privacy-enhancing technology, a lattice-based leveled\footnote{In a ``leveled" FHE scheme, the parameters of the scheme may depend on the depth of the circuits that the scheme can evaluate (but not on their size)~\cite{leveledFHE}. In simpler words, in a leveled FHE, functions are computed only up to a fixed complexity or level. There exists a conversion technique from a leveled FHE to (normal) FHE.} FHE scheme based on the Ring Learning With Errors (RLWE) problem is implemented for the protection of privacy and security of genomic data in i2b2~\cite{homomophicencryption}. The i2b2 is an open-source framework to enable sharing, integration, standardization, and analysis of clinical research data via collaborative efforts. 
	In another work, (leveled) homomorphic encryption was implemented to conduct predictive analyses (e.g., logistic regression) on medical data~\cite{FHEimplementation1} privately in a cloud service. In a recent work, it is used to achieve genome-wide association study, which compares genetic variants and single-nucleotide polymorphisms of genetic data, in a secure and private way~\cite{gwashomomorphic}.

	\subsubsection{Multiparty Computation:} \label{multipertycompt}
	Multiparty Computation (MPC) enables distributed computations on encrypted data without decryption. It eliminates the need for a central trusted party for computations. Each data input is divided into two or more shares and distributes them among the multiple (distrustful) parties. All parties follow a protocol and jointly compute a function on their inputs without revealing their inputs to any other party. The final result is shared among them. In MPC, it is not required to store all data from different parties centrally, for which one needs to have a trusted third party. Yao~\cite{MPC2} first introduced MPC in the early 1980s. It has been an important technique for privacy-preserving computations where data are distributed. The computational models of MPC include boolean, arithmetic, fixed/floating, and random access machine (RAM). For a secure MPC, it is proved that there is a bound on the number of parties being controlled by adversary or colluding~\cite{MPC1}. Homomorphic encryption~\cite{MPC4}, garbled circuits~\cite{MPC2}, linear secret sharing~\cite{MPC3}, and Oblivious Random Access Machine techniques~\cite{ORAM,ORAM1} have been utilized for the construction of secure MPC protocols. MPC has also been used in secure neural network training~\cite{MPCmachinelearning}. 
	
	Unlike HE, MPC has a low computational cost, but it has a considerable communication cost as its processes need to communicate encrypted data with each other across the network, and the communicating parties must remain online during joint computation. In MPC, the correctness of the computation (output) is ensured~\cite{correctnessMPC}. Scalability is another issue with MPC. As the final result (after computation) may leak information about the inputs, MPC alone is not sufficient for privacy. Thus, a combination of MPC with other techniques such as differential privacy (see Section~\ref{sec:diffprivacy}) or secure enclave is required for better privacy results.

	\paragraph{MPC and healthcare:}
	Healthcare is the best use case of MPC. In Scalable Oblivious Data Analytics (SODA)~\cite{SODA}, a Horizon 2020 project, MPC is implemented as an underlying technology to preserve privacy whilst processing personal (health) Big Data from multiple distrusting parties (e.g., hospitals and insurance company). Refer to Section~\ref{sec:soda} for details. 
	In another project, named San-shi, which is a secure computation system developed by Nippon Telegraph and Telephone Corporation (NTT), MPC is implemented for aggregation and statistical processing of confidential data whilst keeping the data encrypted~\cite{sanshi,sanshiusecase}. Refer to Section~\ref{sec:sanshi} for details.  
	In a different work, a privacy-preserving patient linkage technique is developed by using secure MPC based on Sharemind framework~\cite{Sharemind1}. Sharemind~\cite{sharemind} provides a secure infrastructure that hosts (usually) three nodes. Its framework is written in C++. It processes privacy-preserving algorithms, and the security is achieved via secure MPC on additive secret sharing.    
	
	\begin{table}[!t]
		\renewcommand{\arraystretch}{1.05}
		\footnotesize
		\caption{Summary of security and privacy preserving techniques for data-in-use}
		\label{sumtable}
		\centering
		\begin{tabular}{|p{6cm}|c|c|c|c|}
			\hline
			& TEE 		& HE 		& MPC 	& DP	 \\
			\hline
			Interactive 	&No 	&No 	&\textcolor{green!70!blue}{Yes} &No \\
			\hline
			Collaborative
			computing 		&Applicable		& Applicable	& \textcolor{green!70!blue}{Suitable}		& \textcolor{green!70!blue}{Suitable}	\\
			\hline
			Implementation
			complexity (relative) & Low & High 		& High 			&Very Low		\\
			\hline
			Ensures
			correctness 	& No 	  	&  No	&\textcolor{green!70!blue}{Yes}	& No		\\
			\hline
			Computation
			speed (relative) 	&Fast  	& Slow 			& Slow	&Fast		\\
			\hline
			Cryptographic
			technique	 	&No  	& \textcolor{green!70!blue}{Yes} 			& \textcolor{green!70!blue}{Yes}			&No		\\
			\hline
			Output data
			privacy 	&No  	& No 			& No	&\textcolor{green!70!blue}{Yes}		\\
			\hline
			Data
			protection 	& \makecell{Storage \& \\ Computing}  	&\makecell{Storage \& \\ Computing} 			& Computing		& \makecell{Released\\ (output)\\ data}		\\
			\hline
			Mathematical
			guarantee of privacy 	& No  	&No 			& No &\textcolor{green!70!blue}{Yes}		\\
			\hline
			Network Communication &\textcolor{green!70!blue}{Low}	&Not required		&\makecell{High}	&Not required\\
			\hline
		\end{tabular}
	\end{table}
	
	\subsubsection{Differential Privacy:} \label{sec:diffprivacy}
	Differential Privacy (DP) provides the privacy of output data or results from computation or process such that the output data or results only reveal the permitted (which is usually negligible) amount of leakage of an individual input data. Dwork et al.~\cite{differentialprivacy1} introduced DP in 2006. They provided an information-theoretic notion of privacy, called $ \epsilon $-differential privacy, of a randomized algorithm. If the algorithm provides $ \epsilon $-differential privacy for a small (near to zero) $ \epsilon $, then adding or removing one data from its input data set does only nominal change to the outcome of the algorithm (the outcome lies within the multiplicative factor of $ exp(\epsilon)$)~\cite{differentialprivacy1}. In other words, a differentially private output ensures that any participant will not be affected adversely by allowing his/her data for analysis irrespective of studies and available datasets. The most common methods of realizing differentially private algorithms are Laplace mechanism~\cite{differentialprivacy} and exponential mechanism~\cite{differentialprivacy2}, where a random noise generated from Laplace distribution and a scaled symmetric exponential distribution is added to the output data to achieve DP, respectively. The added noise changes the output data nominally, and one can accurately learn the data, but it is sufficient enough to blur the individual input data (which cannot be learned precisely). More noise will make the data more private. Still, it reduces the quality of data and hence its utility. Consequently, a proper trade-off between privacy and utility is always desirable. 
	
	There are two types of DP, namely \emph{local} and \emph{global} differential privacy. In local DP, noise is added by each distributed participant to their input data before collection or computation, whereas in global DP, noise is added to the final output after computation. DP is an important privacy-preserving technique due to its properties. Some important features of DP~\cite{diffprivacybook} are the following:
	\begin{itemize}
		\item DP is immune to post-processing: Any post-processing of the output of a differentially private algorithm cannot make it less differentially private without additional information about the input (private) database.	
		\item Composition of differentially private mechanisms is also differentially private: The composition of differentially private mechanisms are also differential private, where the total privacy losses are cumulative. Thus there can be a significant privacy loss when multiple differentially private computations are performed on an individual's data for a long time.	
		\item The privacy guarantee drops linearly with the size of the group: The privacy guarantee deteriorates with the increase in the group size. Group privacy is different from composition privacy.
	\end{itemize}
	DP assumes that the initial data holders are always trusted, which may not be true in practice. It is a promising privacy-preserving technique but still has limitations, including difficulties in general computing of a global sensitivity that both guarantee privacy and acceptable level of noise and non-compact uncertainty (e.g., Laplace mechanism can change the original answer)~\cite{diffprivacycriticism}. 
	
	\paragraph{Differential privacy and healthcare:}
	DP has been extensively used in the healthcare domain, including releasing health data for research, and their analytic computations. As MPC alone does not guarantee privacy, a combination of DP and MPC is proposed as an underlying technology to preserve privacy whilst processing personal health data in SODA project~\cite{SODA}. DP is also combined with an encryption technique. A combination of encryption with DP is used to guarantee the privacy of genomics data in a distributed clinical setup~\cite{diffprivacy5}. In another work, a DP framework is integrated with the classical statistical hypothesis testing and applied to clinical data mining examples~\cite{diffprivacy4}. DP is implemented in distributed deep learning of two clinical data sets~\cite{diffprivacy6}, where measurements of cumulative privacy loss are done by using Renyi differential privacy~\cite{diffprivacy7}.

	Before discussing the best available techniques and methods through a conceptual system model for the PH data security and privacy, we briefly introduce some important terms and approaches that we are using. Besides, we explore some relevant ML paradigms to healthcare, which considers PH data-in-use, in the following section. 
	\section{Methods of data storage, computing, and learning} \label{sec:computingandlearning}
	
	\subsection{Data storage and computing approaches}
	
	\subsubsection{Data storing methods:}
	There are two common ways of storing PH data, namely \emph{centralized storage}, and \emph{decentralized storage}.  
	In centralized storage, all the data from different sites (e.g., various hospitals) are collected and stored in one central server. It will be easier for computations if all data are available in one server. However, if the server failed or compromised, then this affects all data and systems associated with it. Besides, it is required to trust the server, and this increases the responsibility of the server to protect privacy and maintain the security of the stored information. Some examples of medical projects using centralized storage are 100,000 Genomes Project~\cite{genomeproject} and 23andMe~\cite{23nMe}. In decentralized storage, the data are stored in multiple data servers. For example, each hospital can have its own data storage server. The data servers localize the risk of failure and attacks. However, it is relatively difficult for computations on distributed storage over centralized storage. Some examples of medical projects using decentralized storage are Global Alliance for Genomics and Health (GA4GH)~\cite{GA4GH}, Swiss Personalized Health Network (SPHN)~\cite{SPHN}, and MedCo~\cite{MedCo}.       
	
	\subsubsection{Computing approaches based on data accessibility:}
	There are two types of computing approaches over the stored PH data\footnote{The storage can be centralized or decentralized.}~\cite{ref1} based on data accessibility. These are (1) Data-to-modeler, and (2) Model-to-data.  
	\paragraph{Data-to-modeler:}
	In the Data-to-modeler (DTM) approach, a data modeler has direct access to the data for the model development (training and validation) and hypothesis testing. This approach does not follow the norms of the privacy-by-design approach because the data modeler needs to be trusted, and data, which is sensitive in the healthcare domain, is directly accessed by the modeler. In some cases, this approach is infeasible. For example, different hospitals and insurance companies may not want to share their raw patient data directly with each other due to privacy concerns or competition or legal reasons. Data-to-modeler is a common approach whilst carrying out data analytics. As an example, for research purposes and discoveries, the anonymized health data sets are publicly released, such as health data provided by the Australian Government~\cite{Ausgovhealthdata}, HealthData.gov~\cite{healthdatagov}, and European Data Portal~\cite{europeandataportal}.

	\paragraph{Model-to-data:}
	In Model-to-data (MTD), the data is not directly accessible, and a modeler needs to submit their codes and models (obtained from their initial data) to the data contributor. In other words, the model moves to the data. Afterward, the models are trained and validated at the data contributors' server (or device) on the unseen actual data. The updated model or result is transmitted back to the modeler. The approach of MTD is privacy-by-design and data-centric. This approach seems promising in the case of health data analytics because it enables computing without seeing the patients' personal information. Federated learning (refer to Section~\ref{sec:nopeek} for details) uses the model-to-data approach, and it has been implemented in medical data analytics, including semantic segmentation models on multimodal brain scans~\cite{fedlearningapplication} and prediction of mortality and hospital stay time~\cite{fedlearningapplication2}.

	\subsubsection{Types of computing:}
	There are three types of computing, namely (1) centralized computing, (2) distributed computing, and (3) decentralized computing.  
	
	In centralized computing, all the computations are carried out on one system/server. Use case examples of centralized computing are web application servers and mainframes. The centralized computing has major disadvantages, including single-point failure, scalability issues, and processing speed. On the other hand, in distributed computing, the computations are distributed to multiple systems or servers, but the process control and service requests are handled by one central system/server. Unlike centralized computing, this has no single point failure (increases reliability), scalable, and higher processing speed due to the parallelization of processing over multiple systems/servers. Hadoop~\cite{hadoop}, an open-source software, enables the distributed computing. Servers those running Hadoop, for example, Amazon EC2~\cite{amazonec2}, provide distributed computing services. 
	In decentralized computing, both the computations and control of the processes are distributed among multiple systems/servers. Each computing node can process service requests. An example use case of decentralized computing is Blockchain~\cite{bitcoin}. For example, iExec provides blockchain-based decentralized cloud computing services~\cite{iexec}, and Golem delivers a decentralized marketplace for computing power (anyone can share their unused computing resources)~\cite{golem}. Decentralized computing includes all the benefits of distributed computing. In addition, it offers high availability and autonomy due to multiple service processing nodes. The complexity of the computing environment increases with the increase in the size of the network.
	For health-related data, including genetic data (usually Big Data) processing and analytics, either decentralized or distributed computing is more appropriate over centralized computing. This is because of high computational and storage requirements.

	\subsection{Machine Learning paradigms and healthcare} \label{sec:mlparadigmsandhealthcare}
	In this section, we present some relevant ML paradigms that are relevant to the healthcare domain. Table~\ref{learningtable} provides a summary of these learning paradigms. These paradigms shed light on the PH data use whilst computing, and its transformation to knowledge in the form of ML models. This enables us to distinguish the privacy-preserving ML techniques required for PH data.  
	
	\subsubsection{Transfer Learning:}
	Transfer learning~\cite{transferlearning} leverages pre-trained ML models by reusing them for new related problems. In other words, it transfers the knowledge gained from one problem (in one domain) to the related target problem (in another but similar domain). This transfer of knowledge improves learning in the target problem. The transfer learning approach is useful to train a model even if the data is insufficient because the model is pre-trained with sufficient data on the related problem. It is used in Deep Learning (which requires a large amount of data to model its neural networks)~\cite{deeplearning}, Natural Language Processing (enables machines to understand, process, and manipulate human language) and Computer Vision (allows machines to process images to identify objects). This learning methodology is inappropriate if the problems are not sufficiently related. Transfer learning in deep learning suffers from catastrophic forgetting, meaning that the network forgets its previously learned information once it learns the new information~\cite{catastropiclearning}. Various works, including learning without forgetting~\cite{learnignwithoutforgetting}, progressive neural networks~\cite{progressiveneuralnetwork}, and elastic weight consolidation~\cite{Kirkpatrick3521}, have been proposed to address the catastrophic forgetting problem.   
	
	\paragraph{Transfer Learning and healthcare:} Transfer learning has been extensively used in medical image analysis~\cite{trasferlearningusecase1,TLusecase2}. The learned codebook from 15 million images collected from ImageNet~\cite{imagenet} is used in Otitis Media (a group of inflammatory diseases of the middle ear) images with a detection accuracy of 88.9\%~\cite{trasferlearningusecase1}. In another work, GoogLeNet~\cite{googlenet} and AlexNet~\cite{Alexnet} models are reused and shown to be useful for thoracoabdominal lymph node detection and interstitial lung disease classification problems~\cite{TLusecase2}. A trained model on the ImageNet~\cite{imagenet} database has been reused to train fundus images for the detection of Glaucomatous Optic Neuropathy with higher performance and faster convergence~\cite{deeplearningTL}.         
	
	\subsubsection{Multi-task Learning:}
	Multi-task learning is an approach to inductive transfer that improves generalization by using the domain information contained in the training signals of related tasks as an inductive bias~\cite{multitaskfirstpaper}. It adopts the concept of collective transfer learning, where one task helps another related task to learn better. In other words, multi-task learning aims to improve the overall performance of multiple associated tasks~\cite{surveyonmultitask}. Unlike transfer learning, which focuses on solving one task at a time, multi-task learning solves the various tasks at one time by leveraging the similarity and differences across tasks. As all tasks are learned at the same time, the gained knowledge is available to all tasks. This process of learning is also called parallel transfer. The order in which the tasks are trained makes a difference in transfer learning; on the other hand, in multi-task learning, due to the parallel transfer, it does not make any difference. Thus there is no need to define a training sequence in multi-task learning~\cite{multitaskfirstpaper}.

	\paragraph{Multi-task Learning and healthcare:} 
	Multitask learning has been applied extensively in the healthcare domain. It is used for drug discovery, where models were trained on 259 datasets, including PubChem BioAssay, datasets designed to predict interactions among proteins and small molecules, and database designed to avoid common pitfalls in virtual screening~\cite{multitaskdrugapp}. It is shown to have a significant performance in terms of accuracy to baseline ML methods such as logistic regression and random forest~\cite{multitaskdrugapp}. In another work, a multi-task learning formulation (temporal group lasso multi-task regression) for predicting the Alzheimer disease progression is proposed~\cite{multitaskusecase2}. The disease progression is measured by cognitive scores based on baseline measurement, and the effectiveness of the proposed formulation is evaluated by experimental studies on the Alzheimer disease neuroimaging initiative database. In another work, conditions of mental health based on social media text are modeled as tasks in multi-task learning. This work shows that the model predicts potential suicide attempts with an accuracy of above 80\% in limited training data conditions~\cite{multitaskusecase3}. Besides, applications of multi-task learning include clinical prediction~\cite{multitaskusecase4}, decompensation prediction (predicting whether the patient's health will deteriorate in the next 24 hours)~\cite{multitaskusecase5}, and ECG data analysis~\cite{multitaskusecase6}.

	\subsubsection{Continuous Learning:}
	Continuous learning is a type of ML method that offers to learn from the newly available data over time such that it retains its previously gained knowledge and selectively transfers that knowledge to learn a new task. This way, the model gets benefited from the newly available data without learning from scratch each time whenever data is available. Continuous learning is suitable for the cases where data gradually available over time, i.e., data streaming (e.g., data coming from health monitoring devices), or the size of data is big enough to be out of system's memory, i.e., big data (e.g., human genome information). This learning is usually done for better performance (e.g., accuracy). There are various terms including incremental learning~\cite{incrementallearning,incrementallearning2}, online learning~\cite{misclearning}, and lifelong learning~\cite{lifelonglearning} in literature, and we refer all to the continuous learning as these learning methods learn continuously with time. Continuous learning uses Transfer learning as its integral part. However, transfer learning is not concerned with continuous learning and knowledge retention. The disadvantage of continuous learning includes continuous computation and prone to catastrophic forgetting. 
	
	\paragraph{Continuous Learning and healthcare:}
	Incremental learning is used to build a medical image segmentation framework~\cite{usecaseincrementallearning}. This framework is applied to lung boundary delineation in High Resolution Computed Tomography scans, and experimental results show that it outperforms the fixed or non-adaptive algorithms. In another work, incremental learning algorithms, called $ i^+$Learning and $ i^+$LRA, are introduced based on decision tree learning methodology~\cite{usecaseIL2}. On various medical data sets from the UCI repository~\cite{UCIrepo}, these algorithms are shown to be providing better classification accuracy than the other learning algorithms, including incremental tree induction~\cite{ITIalgo} algorithms.

	%

	\subsubsection{Ensemble Learning:}
	Ensemble Learning implements several (similar or different) learning algorithms (models) to solve the same problem, and then combine the outputs from each algorithm (model) to get a better final result in terms of predictions or classification~\cite{misclearning,ensemblelearning} of the problem. Usually, the final result is decided based on the voting (e.g., majority voting, weighted voting, and Bayesian voting) and statistical process (e.g., averaging and bootstrap aggregating). The voting and statistical methods are usually done for classification and regression, respectively. Support Vector Machine is also used in the voting process~\cite{ensemble2}. We can implement the ensemble learning in two ways. In the first way, different learning algorithms are trained over the same dataset, whereas in a second way, the dataset is split into disjoint subsets, then the same or different learning algorithms are trained over each subset of the dataset. Ensemble learning can be used in incremental learning.
	
	\paragraph{Ensemble Learning and healthcare:} Ensemble learning has been applied in the medical diagnosis of Alzheimer disease based on Magnetic Resonance Imaging datasets~\cite{ensembleusecase1}. Experiments are performed on 416 subjects of the OASIS database implementing Extreme Learning Machine, Bootstrapped Dendritic Computing (BDC), Hybrid Extreme Random Forest, and Random Forest with BDC scoring the highest of 80.8\% accuracy among others~\cite{ensembleusecase1}. In another work, two adaptive distributed privacy-preserving algorithms based on a distributed ensemble strategy are proposed for the health care domain. The algorithms were tested on a Type-2 diabetic electronic health record dataset and showed that the ensemble learning over distributed datasets is better than the learning on each dataset separately~\cite{ensembleusecase2}. In a different work, an ensemble learning, called asBagging\_FSS, is proposed~\cite{ensembleusecase3}. The asBagging\_FSS algorithm is shown to have a good performance (i.e., accuracy) on high dimensional and imbalanced biomedicine datasets, including cancer DNA microarray and cancer protein mass spectrometry datasets~\cite{ensembleusecase3}.  
	
	\begin{table}[!t]
		\renewcommand{\arraystretch}{1.05}
		\footnotesize
		\caption{Summary of machine learning paradigms}
		\label{learningtable}
		\centering
		\begin{tabular}{|c|c|c|c|c|c|}
			\hline
			& Transfer  & Multi-task		& Continuous	& Ensemble 		& No-peek	 \\
			\hline
			Model reuse/share  	&\textcolor{green!70!blue}{Yes}  &\textcolor{green!70!blue}{Yes}    &\textcolor{green!70!blue}{Yes}  	&No  	&No	 \\
			\hline
			Insufficient data 
			&\textcolor{green!70!blue}{Suitable} &\textcolor{green!70!blue}{Suitable}	&\makecell{\textcolor{green!70!blue}{Suitable} provided that\\ data is available with time}	&No	 &\makecell{\textcolor{green!70!blue}{Suitable} provided that data\\ is distributed and collective\\ data is sufficient}	\\
			\hline
			Continuous learning
			&No	 	&No		&\textcolor{green!70!blue}{Yes}	&No	 &No \\
			\hline
			Multiple (different) models
			&No	 	&No		&No		&\textcolor{green!70!blue}{Yes}	 &No \\
			\hline
			\makecell{Model-to-data and\\ distributed computing} &No &No &No  &No &\textcolor{green!70!blue}{Yes}	\\	 
			\hline				
		\end{tabular}
	\end{table}

	\subsubsection{No-peek Learning:}
	\label{sec:nopeek}
	No-peek learning refers to the distributed deep learning techniques that do not require the sharing of raw data available at distributed sources. It is based on collaborative learning and includes federated learning, split learning, and stochastic gradient descent (SGD) based collaborative learning. 
	
	
	
	\paragraph{Federated Learning:} \label{sec:federated learning}
	Federated learning (FL) is a collaborative ML technique~\cite{fed1,fedkonecny2,fed2,fedkonecny3} developed by Google for training models on the training data that are distributed on mobile devices. It moves the computing to the edge devices. It preserves the data privacy at each participating edge devices. In the traditional ML approach, the training data are usually available in one central server (or datacenter), and the training takes place in that server, which is generally a trusted platform and can see data whilst computing. In contrast, FL approach preserves the privacy of data available at the edge devices by performing collaborative learning that never requires the data in the distributed edge devices to be collected out of those devices. More precisely, firstly, the coordinating server trains a global model based on its available data, then it sends the model to a set of devices. Secondly, each of those devices trains the model and computes an update based on the local training data. Thirdly, the updated model parameters are sent to the coordinating server, which aggregates those updates securely by using secure aggregation protocols~\cite{fedprotocol} without learning individual inputs from the devices. Finally, the coordinating server updates its global model based on the new aggregate model parameters. This process is repeated until the model parameters converge. Thus there is a communication cost in FL. TensorFlow Federated framework~\cite{tensorflow} and PySyft~\cite{pysyft} library are enabling the implementation of FL. Besides, FL is an active field of research~\cite{fed3}.      
	
	Challenges in FL includes communication cost between the edge devices and the coordinating server, availability of edge devices, massive distribution of edge devices, and privacy-preserving aggregation at the coordinating server. Besides, attacks such as inference attack (leaking information from the aggregated updated model parameters)~\cite{inferenceattack,modelinversionattack,informationleakage}, and poisoning attack (poison the training data to compromise the global model)~\cite{poisionattack} are possible in FL environment. It is not guaranteed that the updated model parameters to the coordinating server would not reveal sensitive information about the user, so FL requires other privacy-preserving techniques, including Differential Privacy (see Section~\ref{sec:diffprivacy}) for robust privacy protections.

	\paragraph{Federated Learning and healthcare:}
	FL has a vast application in health data analytics. It makes possible to a distributed ML over patient health records privately and without requiring those data out of the hospital's secure data center or user's personal devices. Intel has demonstrated the first real-world medical use case of FL. Intel uses FL to train a deep learning model without sharing medical imaging data among collaborating clinical institutions. They achieve 99\% of the model performance of the same model trained with the traditional (i.e., centralized) data-sharing method~\cite{fedlearningapplication}. In another work, a novel FL model for an optimization of the performance of predicting mortality and hospital stay time is proposed. The experiments in this work show that the proposed model has predictive accuracy close to that of centralized learning~\cite{fedlearningapplication2}. In separate work, FL is used to predict hospitalization for patients with heart diseases using their electronic health records resided in different hospitals or sources~\cite{federatednewnew}. Recently, FL techniques are implemented to analyze the distributed electronic medical records (including non-IID ICU patient data) for an efficient prediction of mortality and hospital stay time~\cite{federatednewpaper}. Federated learning is also employed on brain tumor segmentation data with a comparable segmentation performance to the centralized system~\cite{dfnvidia}.

	\paragraph{Split Learning:}	\label{sec:split learning}
	Split learning, also called split neural network (SplitNN), is a type of distributed deep learning~\cite{splitlearning,splitlearning2,splitlearning3,SplitNN}. Lile FL, SplitNN is useful specially when raw data sharing among the data holders/sources is not possible due to resource limitations or privacy and legal reasons. 
	In this learning, firstly, each client trains a neural network up to a particular layer, called \emph{cut layer}, and transmits the output of that layer, called \emph{activation} or \emph{smashed data}, to the server. After receiving the activation from the client, the rest of the layers of the network is trained by the server and completes the forward propagation. Secondly, the server computes the loss based on the true and the predicted labels, and backpropagate the loss to calculate the correction in the form of gradients until the cut layer. Then the gradient of the cut layer is transmitted back to the client. After receiving the gradients, the client continues the backpropagation of the network on its side. This way of forward and backward propagations continue until the network is trained. The client-side computational and communication requirements are reduced in SplitNN because the client only processes a part of the network, and the data to be transmitted to the server is the activation of the cut layer. The activation is relatively smaller in size compared to the raw data. As the model is split and trained in client and server, SplitNN also protects the trained model's detailed architecture and parameters. This kind of protection is not available in FL. Besides, this technique is similar to FL, as both are distributed ML techniques. SplitNN is shown to have better accuracies and lower computational requirements than FL and large batch synchronous SGD over CIFAR 10 and CIFAR 100 datasets using VGG and Resnet-50 architectures for 100 and 500 clients based setups, respectively~\cite{splitlearning}. However, FL has lesser communication bandwidth than others with smaller (i.e., 100) number of clients~\cite{splitlearning}. It is still an open research problem to make the algorithms and decisions made by deep learning techniques like SplitNN (whose processes are usually blurred to the outside world) explainable as demanded by regulations such as GDPR (EU). Moreover, the explainability factor is associated with the trust of the system, and it is crucial in the health domain.

	\paragraph{Split Learning and healthcare:}
	Configurations of SplitNN, including simple vanilla configuration, U-shaped configuration for split learning without label sharing, and vertically partitioned data for split learning, are proposed for various practical health settings~\cite{splitlearning}. SplitNN is improved to reduce the information leakage at the cut layer of the neural network. The improved SplitNN is called NoPeekNN, and it is tested with a dataset of colorectal histology images without any data augmentation~\cite{splitlearning3}. Recently, SplitNN is studied to analyze the diabetic retinopathy dataset with Resnet-34 and chest X-ray dataset with DenseNet121 in a distributed setting with up to 50 clients. Its accuracy is shown to be better than a non-collaborative setting~\cite{poirot2019split}. In recent works, SplitNN is used together with differential privacy to demonstrate machine learning with privacy on ECG dataset~\cite{split_our1}, and with IoT gateways~\cite{split_our2}.

	\paragraph{Stochastic Gradient Descent:}
	Stochastic Gradient Descent (SGD) based collaborative learning works in similar essence to FL. The difference is that the updates (local updates and global updates) in SGD based collaborative learning are based on one batch of training data, whereas, in FL, a client trains the model for some local epochs before updating it to the server. In SGD based collaborative learning, the parallelization of the updates is done through either model parallelism (a part of the model is evaluated in one machine) or data parallelism (model parameters are computed over a batch of data available in one machine). It includes Distributed Selective SGD~\cite{privacySGD}, Large Minibatch SGD~\cite{LBSGD}, and Distributed Synchronous SGD with backup workers~\cite{synchronousSGD,splitlearning2}.  
	
	In Distributed Selective SGD, each client chooses a fraction of parameters (of its local model) to be updated at each round and share it asynchronously with other clients through a parameter server. On the other hand, Large Minibatch SGD considers synchronous SGD by dividing SGD mini-batches over a pool of parallel workers, and gradient aggregation. The updates are done synchronously, meaning updates on model parameters take place only after receiving all updates from all machines. As the straggling clients slow down the whole process, synchronous optimization with backup workers (computing nodes) in distributed synchronous SGD is proposed~\cite{synchronousSGD}.


	\section{Related health projects, their privacy and security measures} \label{sec:relatedhealthprojects}
	In this section, we present some relevant projects addressing security and privacy in the healthcare domain. These projects include Horizon2020 funded projects, Swiss personalized health network projects, and Nippon Telegraph and Telephone Corporation project. Table~\ref{projecttable} presents a summary of these projects. 
	
	\subsection{Horizon2020}
	Horizon2020 is the biggest EU research and innovation program, with funding from 2014 to 2020~\cite{Horizon2020}. There are several health-related projects under Horizon2020, including My Health My Data, Scalable Oblivious Data Analytics, and KONFIDO. We present the relevant projects in the following. 
	
	\subsubsection{My Health My Data:}
	The project My Health My Data (MHMD)~\cite{MHMD,MHMDnewsletter,MHMDdeliverable} spans from Nov. 2016 to Oct. 2019. The main aim of this project is to introduce a new way of sharing medical information and empowering their primary owner, the patient, using encryption and blockchain technologies. The key proposed elements include the following:
	\begin{itemize}
		\item Blockchain: Blockchain is used for ensuring the lawfulness and legitimacy of data exchange, keeping track of data access, and improving data integrity. MHMD implements blockchain by identifying its key features in a healthcare context, including high transaction rates, low network latency, low energy consumption, scalability, and robust privacy features. More precisely, a hyper ledger is implemented because the healthcare ecosystem requires a federated blockchain with a consensus mechanism such as proof of stake or practical Byzantine fault tolerance. 
		\item Smart contracts: Smart contracts are self-executing protocols that can facilitate, verify, and enforce the performance of the contract. MHMD uses it to automate peer-to-peer transactions under user-defined data access conditions. The MHMD smart contract includes data registration, data access request, privacy-preserving for data transactions, study updates, and an indication of available data matching the study request.
		\item Personal Data Accounts: Personal Data Accounts (PDA) is a personal data storage clouds that enable individuals to access their data from any sources including wearable, clinical data, and social media,  and use them for personal use. This service is provided through digi.me's secure personal data library and consent access process platform in this project.
		\item Dynamic consent: Dynamic consent~\cite{dynamicconsent} enables the possibility for individuals to provide different types of consent according to distinct potential data uses, taking control over the data access by others and its purpose. In MHMD, it is envisioned to use smart contracts for the implementation of dynamic consent.
		\item Secure computation and privacy-preserving mechanisms: MHMD implements both semantic multi-party computation and partial homomorphic encryption techniques to ensure the secure computation of sensitive information. For data privacy of the published datasets are provided through anonymization techniques such as k-anonymity and differential privacy.
	\end{itemize}  
	
	\subsubsection{Scalable Oblivious Data Analytics:} \label{sec:soda}
	The project Scalable Oblivious Data Analytics (SODA)~\cite{SODA1,SODA2,SODA3,SODA4,SODA5} spans from Jan. 2017 to Dec. 2019. The main aim of this project is to enable practical privacy-preserving analytics on (big) data from multiple datasets with healthcare as the first use case. The fundamental techniques include multi-party computation techniques and differential privacy. In their work, they show that the MPC alone does not guarantee privacy and anonymity because the information leakage can occur from the output. Thus, they combine MPC and differential privacy for big data analytics by using MPC framework MPyC, an MPC tool available as a python package designed to support secure $ m $-party computation tolerating a dishonest minority of up to $ t $ passively corrupt parties~\cite{mpyc}, in three-party mode. For the implementation of MPC, SODA provides improvements and expansion of the FRESCO framework\footnote{A FRESCO is a hybrid-mode MPC (software) framework offering a high degree of modularity, parallelization and preprocessing, and it is available as a Java library.}~\cite{FRESCO}, and use them to implement MPC. The upgrades include application interface, SPDZ preprocessing, network improvements, the addition of TinyTables protocol with semi-honest security, and SPDZ2K protocol~\cite{SPDZ}, new library functionality (e.g., comparison protocol and aggregating functionality based on an encryption technique), and MPC based ML. As a part of the project, an MPC-enabled query compiler, called \emph{conclave}, is proposed. Conclave addresses the implementation issue of MPC regarding the domain-related knowledge and scalability issue with big data sizes. More precisely, conclave considers three parties and allows relational queries to be processed among the parties by turning the queries into a combination of local processing steps and secure MPC steps. Other works towards the realization of general-purpose MPC includes efficient secure computations in rings, scaling up MPC to many parties (introduce short keys for secure computation), communication-efficient honest-majority MPC from batch-wise multiplication verification, two-party computation from oblivious linear function evaluation, differentially private logistic regression, and controlling leakage in MPC on non-integer types.

	\subsubsection{KONFIDO:}
	The project KONFIDO~\cite{KONFIDO,Konfido1,konfidoblockchain} spans from Nov. 2016 to Oct. 2019. The main aim of the project is to advance the eHealth technologies for data preservation, data access and modification, data exchange, and interoperability and compliance primarily across European countries. KONFIDO proposes security components for OpenNCP, which is a software implementation of smart open services for European patients (i.e., epSOS) for seamless health data exchange across borders. KONFIDO proposes the use of the following in its system:
	\begin{itemize}
		\item Cryptographic techniques for the security (integration of Intel SGX and fully homomorphic encryption with OpenNCP, respectively),
		\item physical unclonable functions (based on photonic technology) and commercial off-the-shelf CPU technology for a secure data exchange, homomorphic encryption or secure enclaves for secure data processing,
		\item extension of the selected security information and event management (SIEM) to include multiple independent monitoring nodes, 
		\item KONFIDO Secure Identity Across Borders Linked (STORK)~\cite{STORK} compliant eID support for authentication of the operators (developed an eIDAS-compliant eID for OpenNCP), and
		\item blockchain-based log management system for traceability and liability (e.g., informed consent) of health data sharing and data access permission handling.
	\end{itemize}

	\subsection{Swiss personalized health network}
	Swiss personalized health network (SPHN) is a national initiative designed to promote the development of personalized medicine and personalized health in Switzerland, and this project spans from 2017 to 2020~\cite{SPHN}. It envisions for the harmonization of the information system, and data types between the participant hospitals and research institute, and facilitate the nationwide health data exchange for health research. There are several projects under SPHN. We present one project that is related to the security and privacy of health data in the following.  
	
	\subsubsection{MedCo:}
	The project MedCo~\cite{MedCo,diffprivacy5} aims to develop a workable system that provides security and privacy to the sensitive medical data available for research. This system enables investigators to share their medical data in compliance with regulations. MedCo enables the data sharing through a hybrid or ``somewhat" decentralized approach that overcomes the limitations of a centralized system (e.g., single-point failure), and a fully-decentralized system (e.g., need of operational resources and costs on the clinical sites). MedCo distributes trust among local Storage and Processing Units (SPU) and shared SPU. Local SPU is dedicated to the local site, whereas the shared SPUs provides services to the multiple sites (who do not have sufficient resources to maintain its local site). MedCo is built on top of two existing open-source technologies for medical data exploration, viz., i2b2~\cite{i2b2}, and SHRINE~\cite{shrine}. It is tested in a simulated federation of three sites by focusing on a clinical-oncology case with tumor DNA data from The Cancer Genome Atlas. MedCo proposes to use privacy-preserving techniques, including homomorphic encryption, secure distributed protocols, blockchains, and differential privacy. 
	MedCo implements the additively homomorphic encryption (partially homomorphic encryption that satisfies only the addition of ciphertexts where the key is generated collectively). This encryption ensures end-to-end protection by allowing queries and processing on the encrypted data (without decryption). Also, a set of SPUs (more than one SPU) are required for the decryption of the encrypted data, so the system is robust till all the SPUs in the set are not compromised.

	\subsection{Secure Computational System San-Shi~\cite{sanshi,sanshiusecase,sanshipaper1}} \label{sec:sanshi}
	San-Shi is a project of Nippon Telegraph and Telephone Corporation (NTT). It enables aggregation and statistical processing with the functional performance of encrypted data. The data include multi-facility clinical research data. Basic operations include (1) the data operation, e.g., table join (without leaking the join key), (2) filtering by conditions, e.g., NULL filter and data filter with date/strings comparison, (3) aggregation, e.g., frequency table (cross-tabulation and histogram) and quantity table, and (4) statistics, e.g., total sum and variance, t-test, and Kaplan-Meier method.    
	San-Shi enables data integration among different companies and cross-analysis without releasing the individually held data. The key technologies include encryption techniques and multiparty computation based on the secret sharing scheme. San-Shi's system consists of up to four servers, where a client protects data as shares, registers them in each server, requests computation to each server, and collects the results from the servers. San-Shi is tested on the secure and private analysis of genomics data.

	\section{Discussion}
	So far, we have investigated requirements and available techniques for PH data security and privacy. Now in this section, we list out the best available techniques through an illustration of a conceptual system model.
	
	\subsection{A model for computation on distributed PH data} \label{sec:proposedmodel}
	
	
	
\begin{figure}[!ht]
		\centering
		\includegraphics[width=0.8\linewidth]{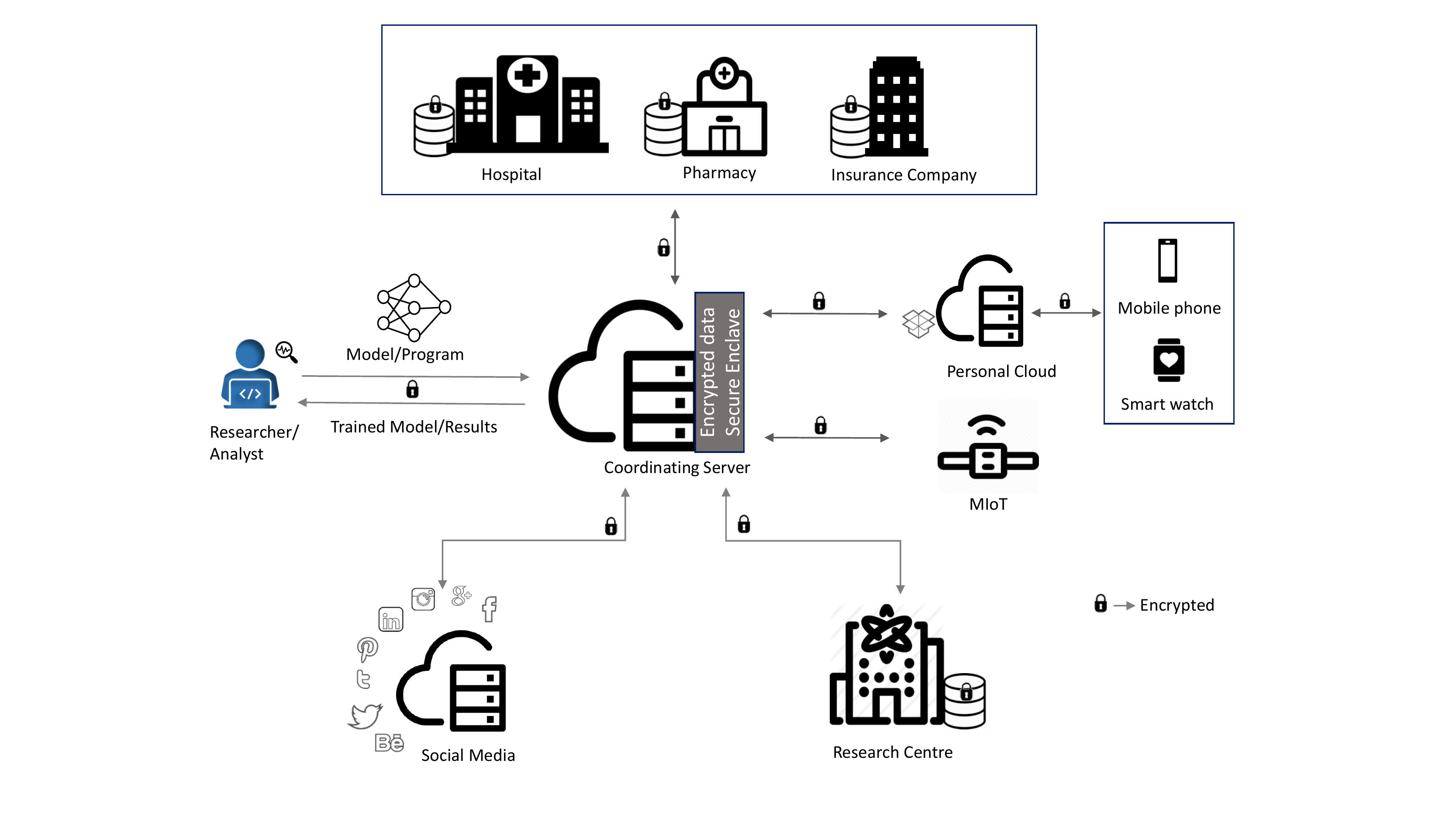} 	
		\caption{An example case of the precision health platform}
		\label{fig:proposedmodel}
	\end{figure}

	
	In a practical setting, health data are decentralized and reside in various locations and institutions, including hospitals, pharmacies, personal mobile devices, research centers, and social media, as shown in Figure~\ref{fig:proposedmodel}. The first challenge is how to ensure the requirements listed in Section~\ref{sec:trustworthyPH}. In this regard, we present three polices and two types of consent. Policies include \emph{consent} policies,\emph{ ethics} policies and \emph{privacy} policies, and consent includes \emph{dynamic consent}, and \emph{smart consent}. 
	
\begin{figure}[!ht]
		\centering
		\begin{subfigure}{0.45\textwidth}
			\hskip-50pt
			\includegraphics[width=1.6\linewidth]{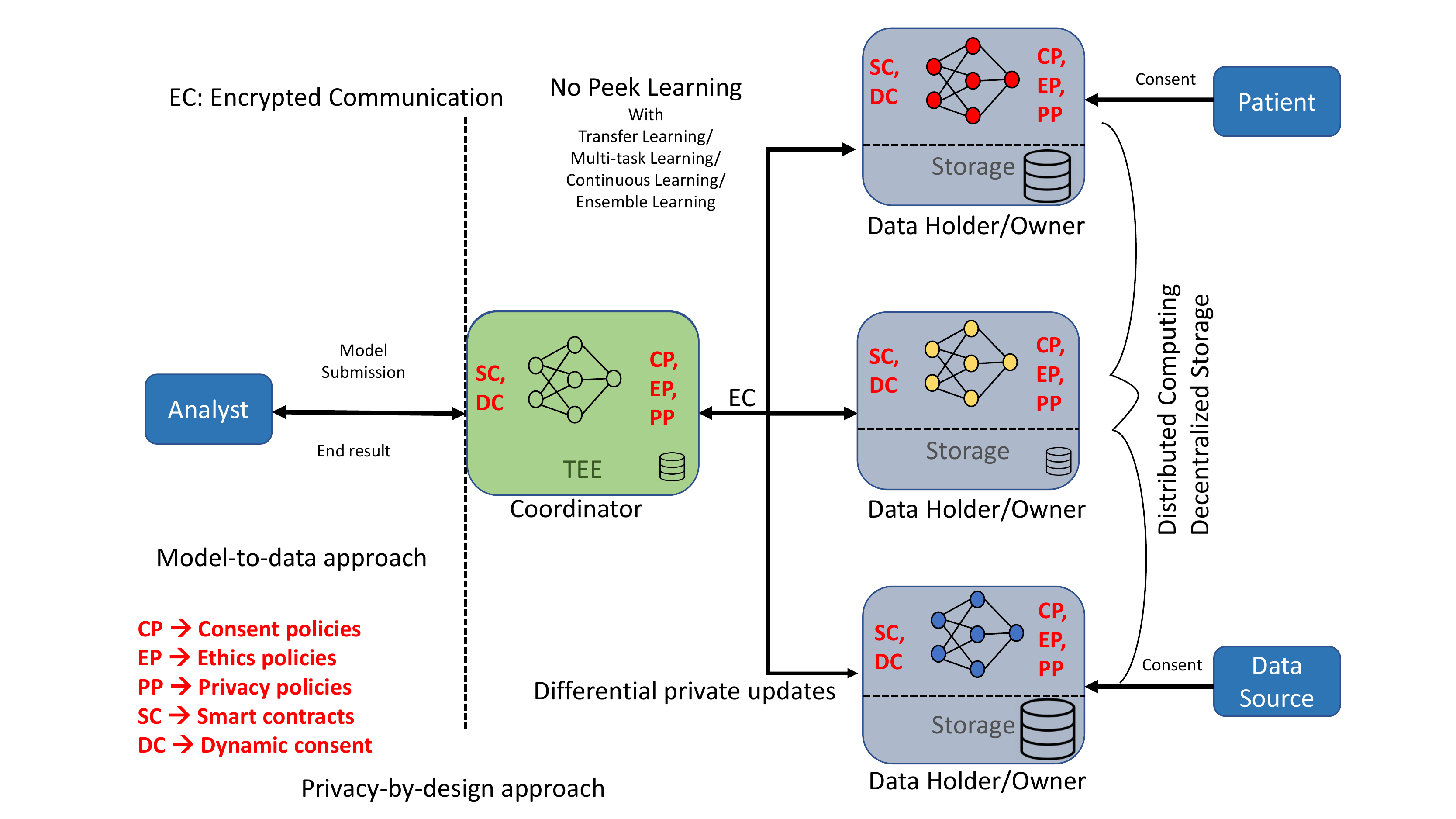}
			
		\end{subfigure}
		\begin{subfigure}{0.45\textwidth}
			\hskip50pt
			\includegraphics[width=0.9\linewidth]{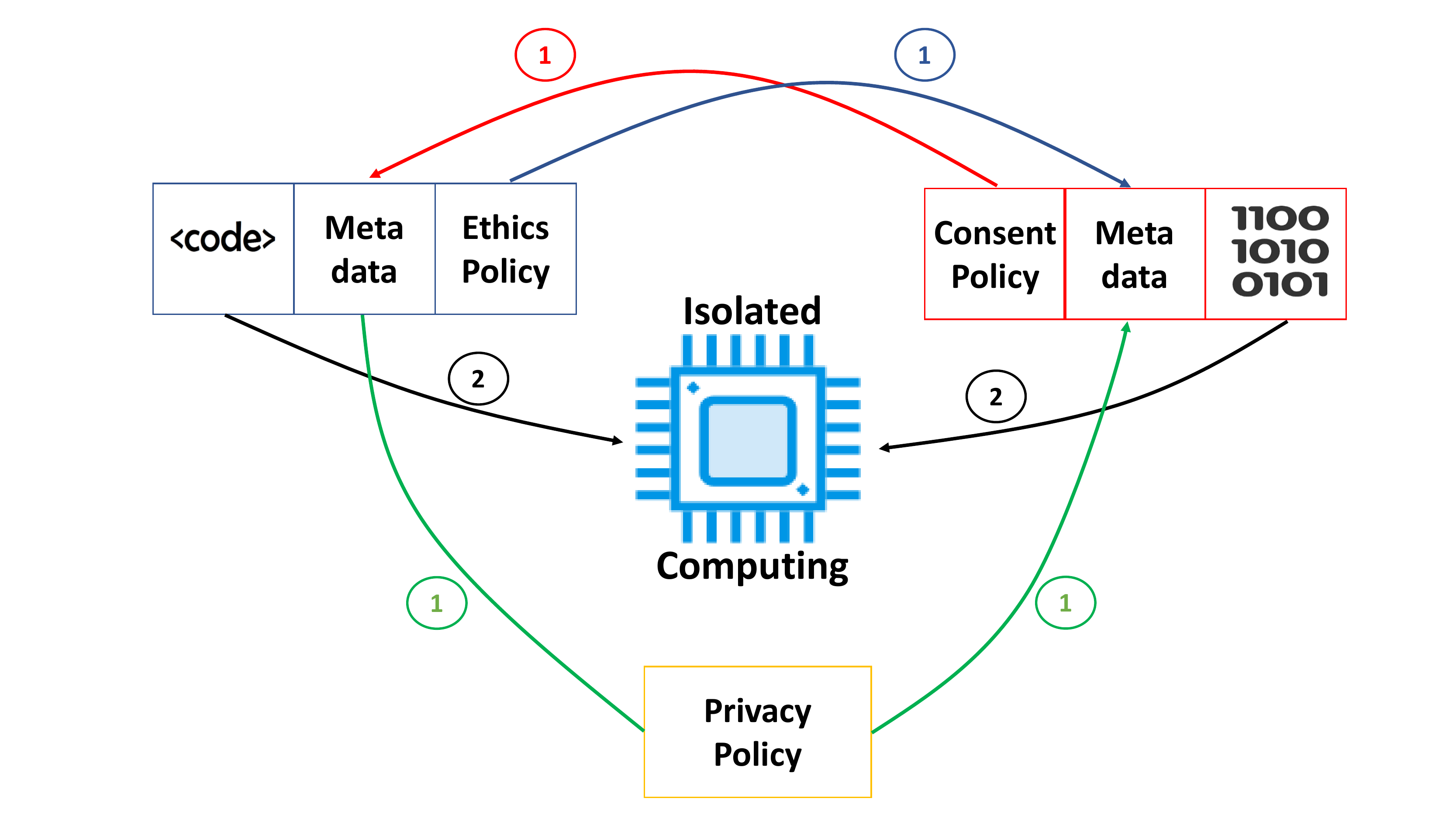}
			
		\end{subfigure}	
		\caption{A model for precision health platform (left figure), and execution of different policies (right figure).}
		\label{fig:proposedmodel1}
	\end{figure}

	Consent policies ensure the fair use of data and check against PH data's metadata. Ethics policies ensure the requirements related to algorithms and analytics, and check against the algorithm's metadata. Privacy policies ensure the overall privacy of the system's environment, including compute, storage and communication, and check against the algorithm's and data's metadata. These policies are the first to be checked before ML meets data (i.e., computation).
	The right figure in Figure~\ref{fig:proposedmodel1} illustrates the execution steps. Overall, each program code (for PH data processing) and PH data has metadata and policy. The policy of one is cross-checked with the metadata of the other, e.g., metadata about data is checked with ethics policy attached to the code, and metadata of the code is checked against consent policy attached to the data. Both metadata are checked against privacy policy as well. These checks are round 1 checks. In round 2, if all checks return true, then code and data can meet for execution. This type of computation assumes the use of cryptography and TEEs. 

	Regarding consent, smart contracts provide a clear description of data handling and its purpose, secure storage of consent, access to withdraw the consent at any time and to renew the consent periodically. On top of it, dynamic consent enables us to have different consents for different uses of the data. How to execute these consent policies and perform their management in a practical healthcare setting is still subjected to further research.     
	
	The other challenge includes health data sharing and integration. It is remained as a constraint to health data analytics because of privacy reasons or reluctant to share the data due to the competition. Thus, the health data have remained in silos. In this regard, we choose an alternative approach to direct data sharing and integration.
	By considering the methods described in Section~\ref{sec:datasecurityandprivacy2} and~\ref{sec:mlparadigmsandhealthcare}, we envision that No-peek learning, specifically, federated learning, split learning and their variant splitfed learning~\cite{split_our3}, make the PH data analytics possible without sharing the data (meaning that the raw PH data never leave their sources such as hospital and pharmacy). Besides, it is not required to rely on other devices or services to delete the participants' sensitive PI once they decide to opt-out from the system (in this regard, it eases the consent management). Thus the No-peek learning maximizes the user's value and increases trust. The model-to-data method, such as No-peek learning, is a privacy-by-design, and it enables distributed computing to handle the distributed PH data. Consequently, medical innovations and development are possible due to the breakage of the data silos (e.g., distributed medical records) in the medical domain. Single security or privacy-preserving technique is not sufficient to provide all requirements, so the presented system model integrates trusted execution environment and differential privacy to the No-peek learning to offer a complete package that ensures the privacy, security, and trust of the system. Other techniques such as homomorphic encryption and multiparty computations can be implemented in the proposed system model, but we exclude them by examining their high computational requirements and their initial stage of technology readiness for a general scenario. Health data security and privacy for data-at-rest and data-in-transit are ensured through the existing methods described in Section~\ref{sec:datasecurityandprivacy1}. Figure~\ref{fig:proposedmodel1} depicts the overview of the platform, and Table~\ref{projecttable} provides a comparative summary with other projects.
	\begin{table}[!ht]
		\footnotesize
		\renewcommand{\arraystretch}{1.05}
		\caption{Summary of the related health projects and their techniques}
		\label{projecttable}
		\centering
		\begin{tabular}{|c|c|c|c|c|c|c|}
			\hline
			Techniques 		& MHMD  & SODA		& KONFIDO	& MedCo 		& San-Shi & Presented model \\
			\hline
			Model-to-data  	&\xmark  	&\xmark   	&\xmark  		&\xmark  			&\xmark  &\textcolor{green!70!blue}{\cmark}\\
			\hline
			Blockchain 		&\textcolor{green!70!blue}{\cmark} 	&\xmark   &\textcolor{green!70!blue}{\cmark}	&\textcolor{green!70!blue}{\cmark}	&\xmark  &Possible	\\
			\hline
			Homomorphic encryption
			&\textcolor{green!70!blue}{\cmark} 	&\xmark	&\textcolor{green!70!blue}{\cmark}	&\textcolor{green!70!blue}{\cmark}	&\xmark	&\xmark\\
			\hline
			Multi-party computation
			&\textcolor{green!70!blue}{\cmark}	&\textcolor{green!70!blue}{\cmark}	&\xmark		&\xmark &\textcolor{green!70!blue}{\cmark} &Possible\\
			\hline
			Differential privacy
			&\textcolor{green!70!blue}{\cmark}	&\textcolor{green!70!blue}{\cmark}	&\xmark			&\textcolor{green!70!blue}{\cmark}		&\xmark &\textcolor{green!70!blue}{\cmark}\\ 
			\hline
			TEE
			&\xmark	&\xmark	 	&\textcolor{green!70!blue}{\cmark} (Intel SGX)		&\xmark	&\xmark &\textcolor{green!70!blue}{\cmark}\\	
			\hline	
			\makecell{Policies enforcer (trustworthy checks)}
			&\textcolor{green!70!blue}{\cmark} &\xmark  &\textcolor{green!70!blue}{\cmark} (Blockchain based) &\xmark &\xmark  &\textcolor{green!70!blue}{\cmark}\\
			\hline
			\makecell{Dynamic and Smart consents} & \textcolor{green!70!blue}{\cmark}  & \xmark & \xmark & \xmark & \xmark & \textcolor{green!70!blue}{\cmark}\\
			\hline
		\end{tabular}
	\end{table}
	
	Now, as an example (refer to Figure~\ref{fig:proposedmodel}), we illustrate the system model considering federated learning, differential privacy, and Intel SGX.
	In this system model, data stored in all end devices and all the communication channels are encrypted at all times, and the authorized recipient having a proper key can only decrypt the data/information. Note that data analytics is an integral part of the precision health system for various purposes, including prediction, diagnosis, and precise treatment of diseases. Now firstly, a researcher/analyst prepares a model based on the available information at his/her ends. Afterward, he/she submits the model to the coordinating server. Secondly, the coordinating server checks the validity of the modeler, including ethics approval at his/her end. Then, the server and data holders execute the policies, dynamic consent, and smart contracts separately, and update their policies as needed. If the executions return true values, then the coordinating server transmits the received model, which is considered as a global model, to all participating data holders/sources (a data holder can be an institution's private server or cloud or a device such as a mobile phone). Thirdly, after receiving the global model, each data holder trains the model on their local datasets and sends the trained model or updates to the coordinating server in an encrypted form. Fourthly, after receiving all the encrypted private updates from the participating data holders, the coordinating server securely aggregates the updates in its secure enclave provided by Intel SGX, and it updates its previous global model.
	Afterward, a differential privacy measure, called global differential privacy, is applied to the updated global model's parameters and send back to the clients. This privacy measure ensures the guaranteed privacy of the local data at some cost of performance from the attacks, including membership inference~\cite{membershipinferenceattack,diffprivacyfedlearning,clientleveldiffprivacy}. This process of federated training takes place continuously until the convergence of model parameters. Finally, a trained global model or result is sent back to the researcher/analyst, who uses this model to analyze the health condition of its data subjects. Besides, the individual participants can utilize the final global model for their health analytics. In the above process, the clients can take differential privacy measures to their individual updates to the coordinating server. This is known as local differential privacy measures. However, the coordinating server's operations are confined within the trusted platform intelSGX (i.e., a trusted platform), the local differential privacy is not required. Due to composition property, it reduces the overall accuracy than by using global differential privacy because the server computes on the noisy (differentially private) local updates.

	There are various challenges in the implementation of the presented system. Some of the difficulties are due to the participants with different computing power, low computing speed, dropped connections, data heterogeneity, high communication cost, and malicious participation. Firstly, splitting the model for its distributed training, such as in SplitNN~\cite{splitlearning} or a variant of SplitNN~\cite{split_our3}, is a possible solution for participants with low computing power. This is because each participant is responsible for training only a part of the model. However, we may require other models besides neural networks in the platform. Secondly, regarding low computing participants, the addition of trusted workers (like in the case of large batch synchronous SGD), who compute and process on behalf of the participant, in the network can be a solution. This way, we can reduce the reporting time of stragglers to the coordinating server. The statistical challenges, including unbalanced data (non-IID data and variation in data points with time) on each participant, and systems challenges, including high communication cost and issues such as stragglers, are addressed through federated multi-task learning framework, e.g., MOCHA~\cite{federatedmultitasking}. Besides, asynchronous updating methods address the problems due to stragglers in the network. Thirdly, one can select a sufficient number of participants to avoid the dropped connections. Lastly, how to prevent malicious participation effectively in the federated setting is still an open problem. However, a strengthened auditing system (which performs various activities including accuracy auditing and log checking) and adversarial training~\cite{adversarialtraining} enable the anomaly detection during model training and secure aggregation.  
	
	\section{Conclusion}
	Precision health (PH) data security, privacy, and trust are of utmost importance due to ethical and regulatory requirements, and they are equally demanded by the data owners/subjects. A detailed requirement for the data security and privacy enable to studying and implementing compliance-friendly techniques whilst handling PH data. In this regard, this paper investigated both the regulations (e.g., HIPAA USA, GDPR EU, and Privacy Act Australia) and ethical guidelines to present a detailed requirement for PH data security and privacy. Our studies observed proper informed consent, secure data transfer, secure and privacy-preserving data processing (computation), proper administrative, physical and technical safeguards, confidentiality, integrity checks, transparency, fairness, availability, minimum and limited use, and breach notification as major requirements from regulatory perspectives. From ethical aspects, studies observed ethics approval, awareness, control, maintaining ownership, limiting information leakage, proper data-sharing management, risk analysis and management along with requirements from regulations, including privacy, security, confidentiality, trust, and breach notification, as essential requirements. Besides, considering the sensitivity of PH data usage in health decision making, health data trustworthiness is a particular requirement. This requires PH data in a standard format, simple, clear, complete, accurate, timely, and transparent for its effectiveness and correctness whilst making inferences for decisions. We analyzed the challenges based on these requirements, and identified four significant challenges; they are health data security and privacy whilst computing, consent management, PH data trustworthiness, and legal and ethical compliance.
	
	Data analytic is an integral part of precision health. This field is evolving with privacy-preserving techniques that are recently introduced and not discussed jointly in the literature. So, we focused our studies on the relevant techniques and recent developments, including health projects, to address PH data security and privacy whilst computing. Our survey found that there are no significant difficulties for the security and privacy of PH data-at-rest and data-in-transit; however, privacy and security of PH data-in-use are still an active research field. In this aspect, we presented and discussed hardware-based techniques, i.e., trusted execution environment, cryptography-based techniques such as homomorphic encryption and multi-party computation, and differential privacy. By providing an overview of machine learning paradigms in healthcare, our studies suggested No-peek learning as a suitable data-centric and privacy-preserving ML computing technique for PH data. This is due to its model-to-data computing approach on distributed PH data (remained in silos). 
	
	To provide an overview and practicality of the techniques that we discussed, we reviewed some notable health projects and their techniques for the security and privacy of health data. Our studies observed that blockchain, cryptography, and hardware-based techniques, differential privacy, and policies for consent and privacy are most commonly implemented. Together with the model-to-data approach, we illustrated an example case for the security and privacy of PH data along with a policy enforcer in a PH platform. Clearly, these techniques enable the trustworthiness and facilitate ethical clearance, consent management, medical innovations, and developments by handling PH data with compliance. As security threats and maturity of the techniques evolve with time, a study of techniques for security and privacy of PH data-in-use is still an active field of research.


	\bibliographystyle{unsrt}  
	\bibliography{cas-refs}


	
\end{document}